\documentclass[aps,prd,twocolumn,showpacs,nofootinbib,floatfix]{revtex4-1}  
\usepackage{graphicx}  
\usepackage{dcolumn}   
\usepackage{bm}        
\usepackage{amssymb}   
\usepackage{atlasphysics}
\usepackage{lineno}
\usepackage{hyperref} 
\usepackage{amsmath}  
\usepackage{lldefs}
\usepackage{multirow}   
\usepackage{rotating}   
\usepackage{longtable}  
\usepackage{afterpage}  

\newcommand{\LumiElectronsInFb}{20.3~fb$^{-1}$}
\newcommand{\LumiMuonsInFb}{20.5~fb$^{-1}$}

\newcommand{\PvalueElectron}{27}%
\newcommand{\PvalueMuon}{28}%

\newcommand{\LimitElectronExpected}{2.76}
\newcommand{\LimitElectron}{2.79}

\newcommand{\LimitMuonExpected}{2.53}
\newcommand{\LimitMuon}{2.53}

\newcommand{\LimitCombinedExpected}{2.87}
\newcommand{\LimitCombined}{2.90}

\newcommand{\LimitCombinedPsi}{2.51}

\newcommand{\LimitCombinedChi}{2.62}

\newcommand{\LimitCombinedExpectedZs}{2.82}
\newcommand{\LimitCombinedZs}{2.85}

\newcommand{\LimitCombinedExpectedG}{2.67}
\newcommand{\LimitCombinedG}{2.68}

\newcommand{\LimitCombinedExpectedGTwo}{3.05}
\newcommand{\LimitCombinedGTwo}{3.05}
	
\newcommand{\LimitCombinedGOne}{1.25}
\newcommand{\LimitCombinedGThree}{1.96}
\newcommand{\LimitCombinedGFive}{2.28}

\newcommand{\LimitCombinedGExpOne}{1.28}
\newcommand{\LimitCombinedGExpThree}{1.95}
\newcommand{\LimitCombinedGExpFive}{2.25}

\newcommand{\LimCombExpMWTCOneFive}{2.24}
\newcommand{\LimCombMWTCOneFive}{2.27}

\newcommand{\LimCombExpMWTCTwo}{1.96}
\newcommand{\LimCombMWTCTwo}{1.99}

\newcommand{\LimCombExpMWTCThree}{1.54}
\newcommand{\LimCombMWTCThree}{1.57}

\newcommand{\LimCombExpMWTCFour}{0.90}
\newcommand{\LimCombMWTCFour}{0.89}

\newcommand{\LimCombExpMWTCFive}{0.56}
\newcommand{\LimCombMWTCFive}{0.57}

\newcommand{\LimCombExpMWTCSix}{0.33}
\newcommand{\LimCombMWTCSix}{0.33}

\newcommand{\LimCombExpMWTCSeven}{0.24}
\newcommand{\LimCombMWTCSeven}{0.24}

\newcommand{\LimCombExpMWTCEight}{0.22}
\newcommand{\LimCombMWTCEight}{0.22}

\usepackage{preprintcover}  
\PreprintCoverPaperTitle{Search for high-mass dilepton resonances in $pp$ collisions at $\rts = 8$~TeV with the \mbox{ATLAS} detector}  
\PreprintIdNumber{CERN-PH-EP-2014-053}  
\PreprintCoverAbstract{The \mbox{ATLAS} detector at the Large Hadron Collider is used to search for high-mass resonances 
decaying to dielectron or dimuon final states.
Results are presented from an analysis of proton--proton ($pp$) collisions at a center-of-mass energy of 8~TeV
corresponding to an integrated luminosity of \LumiElectronsInFb\ in the dielectron channel and \LumiMuonsInFb\ in the dimuon channel.
A narrow resonance with Standard Model \z\ couplings to fermions is excluded at 95\%~confidence level for masses
less than \LimitElectron~TeV in the dielectron channel, \LimitMuon~TeV in the dimuon channel, and 
\LimitCombined~TeV in the two channels combined.  Limits on other model interpretations
are also presented, including a grand-unification model based on the \esix~gauge group, \zstar\ bosons, \MM, 
a spin-2 graviton excitation from Randall--Sundrum models, quantum black holes and a Minimal Walking Technicolor 
model with a composite Higgs boson.}  
\PreprintJournalName{Phys. Rev. D}  

\begin{document}


\title{Search for high-mass dilepton resonances in $pp$ collisions at $\rts = 8$~TeV with the \mbox{ATLAS} detector}  
\author{The \mbox{ATLAS} Collaboration}

\begin{abstract}
The \mbox{ATLAS} detector at the Large Hadron Collider is used to search for high-mass resonances 
decaying to dielectron or dimuon final states.
Results are presented from an analysis of proton--proton ($pp$) collisions at a center-of-mass energy of 8~TeV
corresponding to an integrated luminosity of \LumiElectronsInFb\ in the dielectron channel and \LumiMuonsInFb\ in the dimuon channel.
A narrow resonance with Standard Model \z\ couplings to fermions is excluded at 95\%~confidence level for masses
less than \LimitElectron~TeV in the dielectron channel, \LimitMuon~TeV in the dimuon channel, and 
\LimitCombined~TeV in the two channels combined.  Limits on other model interpretations
are also presented, including a grand-unification model based on the \esix~gauge group, \zstar\ bosons, \MM, 
a spin-2 graviton excitation from Randall--Sundrum models, quantum black holes and a Minimal Walking Technicolor 
model with a composite Higgs boson.
\end{abstract}

\maketitle




\section{\label{sec:intro}Introduction}

The current energy frontier can be explored in the invariant mass spectrum of dielectron or dimuon pairs via a search for new massive resonances at the Large Hadron Collider (LHC).  
Such a search has been performed using the full 8~\TeV\ center-of-mass energy proton--proton ($pp$) collision dataset of about 20 fb$^{-1}$ recorded with the \mbox{ATLAS} detector~\cite{atlas:detector} in 2012.  

While the Standard Model (SM) has been confirmed at the LHC, the identification of massive dilepton resonances 
in proton-proton collisions still constitutes one of the most promising channels in searches for new physics.  
It implies a fully reconstructed signal over a smooth and well-understood background. Models with dilepton 
resonances are predicted in many scenarios for new physics.
Among these are grand-unification models, which are motivated by gauge unification or a restoration of the left-right symmetry violated by the weak interaction.  These models predict the existence of additional neutral, spin-1 vector gauge bosons, called \zp\ bosons, due to the existence of larger symmetry groups that break to yield the SM gauge group and additional $U(1)$ gauge groups.  Examples considered in this article include the \zp\ bosons of the \esix-motivated~\cite{London:1986dk,Langacker:2008yv} and Minimal Models~\cite{Villadoro}.  Another \zp\ signal, the \zpssm, is considered due to its inherent simplicity and usefulness as a benchmark model.  The Sequential Standard Model (SSM) includes a \zpssm\ boson with couplings to fermions equivalent to those of the SM $Z$ boson.

Dilepton resonances are also predicted by several models motivated by solutions to the hierarchy problem of the SM, 
which involves the need to reconcile the very different scales of electroweak symmetry breaking and the gravitational Planck scale (\mpl). 
The search for physics beyond the SM remains as crucial as it was prior to the discovery of the Higgs boson at the LHC~\cite{atlasHiggs,cmsHiggs}, since solving the hierarchy problem is one of the primary objectives of the LHC physics program.  Examples of potential signals in models that address the hierarchy problem are the \zstar~\cite{wzstar,wzstar_motivate,wzstar_refmod,wzstar_refmod2} boson, the spin-2 graviton excitation in Randall--Sundrum (RS) models~\cite{RS}, quantum black holes (QBHs)~\cite{ref:Meade_QBH}, and technimesons in Minimal Walking Technicolor (MWT)~\cite{TC3,TC4,TC5,Foadi:2012bb}.  These, along with the \zp\ interpretations motivated by grand unification, are further discussed in Sec.~\ref{sec:ModelsSection}.  

To conduct the search, the dilepton invariant mass (\mll) line shape is examined for a localized excess of events
corresponding to a new resonance, where $\ell \ell$ corresponds to either the dielectron or dimuon final state. 
This is done using signal and background templates that provide the expected yield of events in bins of \mll.  The
methodology is fully described in Sec.~\ref{sec:stats}.  This search approach is advantageous because using the
full shape of the distribution makes the analysis robust against uncertainties in the background model at high
mass.  If shape information were not used in the analysis, uncertainty in the background estimate would be more
likely to mask a potential signal.  The shape-based method is also more sensitive to a signal in the case of a
signal with a low-mass tail arising from off-shell production, which occurs due to the steeply falling parton
distribution function (PDF) of the two colliding partons at large values of Bjorken $x$.  This feature is commonly
referred to as a ``parton-luminosity tail," and its size increases with the resonance width.  The impact of this parton-luminosity tail on the \mll\ distribution grows as the kinematic limit is approached.

The models considered here predict resonances that are narrow relative to the detector resolution.  In such cases, interference effects, where they occur, are not expected to significantly alter the line shape and are thus not considered.  The exception to this is the class of \MM\ described in Sec.~\ref{sec:ModelsSection}, for which large coupling strengths, and hence larger widths, are considered.  In this case, interference effects are included explicitly in the analysis.

The potential signals studied in this analysis vary in width and spin, and some exhibit a parton-luminosity tail while others do not.  Because of this, the final results given in Sec.~\ref{results} can be interpreted in the context of other models that are not directly studied here, but that predict resonances in the \mll\ spectrum with similar signal shapes.


\section{\label{sec:ModelsSection}Description and status of theoretical models}

A detailed description of the models studied in this article is given in this section. For most models, the best previous limits from the \mbox{ATLAS} experiment were obtained using 5~fb$^{-1}$ of data at $\sqrt{s} = 7$~TeV~\cite{zprime_2011_paper}, while the exclusion results from the CMS experiment are based on 5~fb$^{-1}$ of data at $\sqrt{s} = 7$~TeV and 4~fb$^{-1}$ of data at $\sqrt{s} = 8$~TeV~\cite{Chatrchyan201363}.  The data collected at 7 TeV have not been used to obtain the results presented in this paper, as doing so would not significantly extend the sensitivity of the search.  Previous limits on the mass scale for QBH production are obtained from other sources, as noted in Sec.~\ref{sec:qbh_section}.  

For the benchmark model, previous results from \mbox{ATLAS} exclude a \zpssm\ boson with mass less than 2.22~TeV at 95\%
confidence level (CL), while previous results from the CMS experiment exclude a \zpssm\ boson with mass less than 2.59~TeV at 95\%~CL.  Direct searches at the Tevatron experiments~\cite{Abazov:2010ti,CDF:Zpmumu} and indirect constraints from LEP~\cite{Abbiendi:2003dh,Abdallah:2005ph,Achard:2005nb,Schael:2006wu} have resulted in limits on the \zpssm\ mass of 1.071~\TeV~\cite{CDF:Zpmumu} and 1.787~\TeV~\cite{Langacker:2009su}, respectively.


\subsection{\label{sec:esix_section}\esix-motivated \zp\ models}

In the class of models based on the \esix\ gauge group, this unified symmetry group can break to the SM in a number of different ways~\cite{London:1986dk}.  In many of them, \esix\ is first broken to $SO(10) \times U(1)_\psi$, with $SO(10)$ then breaking either to $SU(4) \times SU(2)_\mathrm{L} \times SU(2)_\mathrm{R}$ or $SU(5) \times U(1)_\chi$.
In the first of these two possibilities, a \zpthreeR\ coming from $SU(2)_\mathrm{R}$ or a \zpBL\ from the breaking of $SU(4)$ into $SU(3)_\mathrm{C} \times U(1)_\mathrm{B-L}$ could exist at the TeV scale.  Both of these \zp\ bosons appear in the \MM\ discussed in the next section.

In the $SU(5)$ case, the presence of $U(1)_\psi$ and $U(1)_\chi$ symmetries implies the existence of associated gauge bosons \zppsi\ and \zpchi\ that can mix.  When $SU(5)$ is broken down to the SM, one of the $U(1)$ can remain unbroken down to intermediate energy scales~\cite{London:1986dk,Langacker:2008yv}. Therefore, the precise model is governed by a mixing angle $\te6$, with the new potentially observable \zp\ boson defined by $\zp(\te6)=\zppsi\cos\te6+\zpchi\sin\te6$.  
The value of $\te6$ specifies the \zp\ boson's coupling strength to SM fermions as well as its intrinsic width.  In comparison to the benchmark \zpssm, which has a width of approximately 3\% of its mass, the \esix\ Models predict narrower \zp\ signals.  The \zppsi\ considered here has a width of 0.5\% of its mass, and the \zpchi\ has a width of 1.2\% of its mass~\cite{Dittmar:2003ir,Accomando:2010fz}.  All other \zp\ signals in this model are defined by specific values of $\te6$ ranging from 0 to $\pi$, and have widths between those of the \zppsi\ and \zpchi.

Previous results from \mbox{ATLAS} exclude the \zppsi\ (\zpchi) boson with mass less than 1.79~TeV (1.97~TeV) at 95\%~CL~\cite{zprime_2011_paper},
while the CMS experiment excludes a \zppsi\ boson with mass less than 2.26~TeV at 95\%~CL~\cite{Chatrchyan201363}.


\subsection{\label{sec:min_models_section}Minimal \zp\ models}
\label{introMM}

In the \MM~\cite{Villadoro}, the phenomenology of \zp\ boson production and decay is characterized by three parameters: two effective coupling constants, \gbl\ and \gy, and the \zp\ boson mass.  
This parameterization encompasses \zp\ bosons from many models, including the \zpchi\ belonging to the \esix-motivated Model of the previous section, the \zpthreeR\ in a left-right symmetric model~\cite{Senjanovic:1975rk,Mohapatra:1974hk} and the \zpBL\ of the pure ($\mathrm{B-L}$) Model~\cite{Basso:2008iv}, where B~(L) is the baryon (lepton) number and ($\mathrm{B-L}$) is the conserved quantum number.  The coupling parameter \gbl\ defines the coupling of a new \zp\ boson to the ($\mathrm{B-L}$) current, while the \gy\ parameter represents the coupling to the weak hypercharge Y.  It is convenient to refer to the ratios $\gbltilde \equiv \gbl / g_Z$ and $\gytilde \equiv \gy / g_Z$, where $g_Z$ is the coupling of the SM $Z$ boson defined by $g_Z = 2 M_Z / v$.  Here $v = 246$~GeV is the SM Higgs vacuum expectation value.  To simplify further, \gammap~and~\thetamin\ are chosen as independent parameters with the following definitions: $\gbltilde=\gammap \cos\thetamin$, $\gytilde=\gammap \sin\thetamin$.  The \gammap\ parameter measures the strength of the \zp\ boson coupling relative to that of the SM $Z$ boson, while \thetamin\ determines the mixing between the generators of the ($\mathrm{B-L}$) and the weak hypercharge Y gauge groups.  Specific values of \gammap\ and \thetamin\ correspond to \zp\ bosons in various models, as is shown in Table~\ref{tab:minimal_models} for the three cases mentioned in this section.

\begin{table}[h]
\caption{
Values for \gammap\ and \thetamin\ in the \MM\ corresponding to three specific \zp\ bosons: \zpBL, \zpchi\ and \zpthreeR.
The SM weak mixing angle is denoted by $\theta_W$.
}
\label{tab:minimal_models}
\begin{center}
\begin{tabular}{l|ccc} 
\hline
\hline
 & \zpBL & \zpchi & \zpthreeR  \\
\hline
\gammap & $\sqrt{\frac{5}{8}} \sin{\theta_W}$ & $\sqrt{\frac{41}{24}} \sin{\theta_W}$ & $\frac{5}{\sqrt{12}} \sin{\theta_W}$ \\
$\cos\thetamin$ & 1 & $\sqrt{\frac{25}{41}}$ & $\frac{1}{\sqrt{5}}$ \\
$\sin\thetamin$ & 0 & $-\sqrt{\frac{16}{41}}$ & $-\frac{2}{\sqrt{5}}$ \\
\hline
\hline
\end{tabular}
\end{center}
\end{table}

For the \MM, the
width depends on \gammap\ and \thetamin, and the interference with the SM \dy\
process is included.  Couplings to hypothetical right-handed neutrinos and to $W$~boson pairs are not included. 
Previous limits on the \zp\ mass versus couplings in the context of these models were set by the \mbox{ATLAS}
experiment; the specific mass limit varies with \gammap.  For $\gammap=0.2$, the range of \zp\ mass limits at
95\%~CL corresponding to $\thetamin \in [0,\pi]$ is 1.11~\TeV\ to 2.10~\TeV~\cite{zprime_2011_paper}.


\subsection{\label{sec:zstar_section}\zstar\ bosons}

One set of models proposes a solution to the SM hierarchy problem via the introduction of a new doublet of vector bosons: $(\zstar, \wstar)$~\cite{wzstar,wzstar_motivate,wzstar_refmod,wzstar_refmod2}.  These are predicted to have masses near the weak scale, motivating the search at the LHC.  

As a result of the tensor form of the coupling, the kinematics of the \zstar\ boson's decay to dileptons are different from that of a \zp\ boson~\cite{wzstar}, and there is no interference between this and the \dy\ process.
To fix the \zstar\ boson's coupling strength to fermions, a model with quark--lepton universality is adopted~\cite{wzstar_refmod,wzstar_refmod2}.  The gauge coupling is chosen to be the same as in the SM $SU(2)$ group, and the scale of new physics is proportional to the mass of the new heavy boson.  
The model parameters are chosen such that the total and partial decay widths of the \wstar\ are the same as those of the charged 
partner of the \zpssm\ boson (\wpssm) with the same mass.  The width of the \zstar\ resonance is 3.4\% of its mass~\cite{wzstar_refmod2}.  

Previous \mbox{ATLAS} results exclude a \zstar\ with mass less than 2.20~TeV at 95\%~CL~\cite{zprime_2011_paper}.


\subsection{\label{sec:graviton_section}Graviton excitations in Randall--Sundrum models}

Models with extra dimensions offer an alternative solution to the mass hierarchy problem in that the higher-dimensional 
Planck scale can be of the order of the electroweak scale.
Among them, the Randall--Sundrum model~\cite{RS} postulates the existence of one warped extra dimension.  
Specifically, the geometry of the original RS model contains two 4-dimensional branes, known as the TeV brane and the Planck brane, within a 5-dimensional bulk.  
The extra dimension in the bulk is compactified, which leads to a Kaluza--Klein tower of excited states of the graviton. 
The particles of the SM are confined to the TeV brane, where due to warping the apparent strength of gravity is exponentially suppressed. 
Gravity originates on the Planck brane; gravitons are also located on the Planck brane, but can propagate in the bulk. 

The RS model phenomenology is characterized by the mass of the lightest Kaluza--Klein excitation mode of the graviton, known
as \gstar, and the ratio \kovermb, which defines the coupling strength of the \gstar\ to SM particles.  Here $k$ is a scale
that defines the warp factor of the extra dimension and $\mplb =\mpl /\sqrt{8\pi}$ is the reduced Planck mass.  The \gstar\
in this model is expected to be narrow for values of $\kovermb < 0.2$.  The intrinsic width of the particle is proportional
to $(\kovermb)^2$, and is 0.014\% (5.8\%) of the pole mass for $\kovermb = 0.01~(0.2)$.  A lower bound on \kovermb\ of 0.01 is theoretically
preferred~\cite{Davoudiasl2000}, as it limits the new physics energy scale to be of the order of TeV, and less than 10~TeV.  
For values above \kovermb\ $\approx 0.1$ the 
compactification radius approaches the Planck length
and is less motivated on theoretical grounds~\cite{Davoudiasl2000},
as this theory does not incorporate quantum gravity.

The \gstar\ is produced predominantly via quark--antiquark annihilation and gluon fusion, with decays to SM fermions or bosons.  
While the branching ratio to dileptons is low due to the spin-2 quantum numbers of the particle, the dilepton final state is nevertheless sensitive to new 
spin-2 resonances due to the clean final state.

Previous \mbox{ATLAS} results exclude a \gstar\ with coupling \kovermb= 0.1 at 95\%~CL for masses less than 2.16~TeV~\cite{zprime_2011_paper},
and the corresponding limit from CMS is 2.39~TeV~\cite{Chatrchyan201363}.


\subsection{\label{sec:qbh_section}Quantum black holes}

In the context of models with extra dimensions, semi-classical black holes can be formed at a collider if the available energy is well 
above the higher-dimensional Planck scale~\cite{ref:Dimopoulos_QBH,ref:Giddings_QBH}. Such black holes would then decay through Hawking radiation. 
Quantum (or non-thermal) black holes differ from these variants in that they lack a well-defined temperature or significant entropy. 
This inhibits thermal decays of black holes produced at a mass scale just above the (higher-dimensional) Planck scale, 
which in turn limits the number of particles in the final state~\cite{ref:Meade_QBH}.  
For two-particle final states, it is interesting to look at the quantum gravity regime, where the threshold for QBH production, $M_{\mathrm{th}}$, 
lies between the higher-dimensional Planck scale, and about five times this value~\cite{ref:Meade_QBH,ref:GingrichMartell_QBH,ref:Calmet_QBH}.  
The QBH decay is governed by the yet unknown theory of quantum gravity, but it is assumed that QBHs 
emit with equal strength all SM particle degrees of freedom.
Provided the higher-dimensional Planck scale is not higher than a few TeV, QBHs could be observed at the LHC.

Production of QBHs can occur in the original RS model, and in the extra-dimensional model proposed by Arkani-Hamed, Dimopoulos and Dvali (ADD) \cite{ArkaniHamed:1998rs}. 
Both scenarios are considered in the model interpretation presented here. The ADD model postulates the existence of $n \geq 1$ flat additional spatial dimensions, 
commonly compactified with radius $R$.  Only gravity propagates in these extra dimensions, with SM particles confined to a 4-dimensional manifold.  
The threshold for QBH production in the ADD model is assumed to 
correspond to the higher-dimensional Planck scale. 
The analysis here was performed assuming $n=6$, 
but the dependence of the resulting production limit on $n$ is small.

The specific model~\cite{ref:Gingrich_QBH} used to interpret the result of this article conserves color, electric charge and total angular momentum.  
Two QBHs states with zero charge, produced via $q\overline{q}$ and $gg$, have predicted branching ratios to each dilepton final state
of 0.5\% and 0.2\%, respectively, assuming conservation of the global symmetries of lepton and baryon number. 
While the model parameters of Ref.~\cite{ref:Gingrich_QBH} are considered in the context of ADD, one can take the 5-dimensional ADD case to obtain 
an approximate RS model, which is what is used in the case of the RS model interpretation.
In the RS model, the higher-dimensional Planck scale
$\tilde{M}$ can be calculated from the \gstar\ mass and \kovermb\ as follows~\cite{ref:Meade_QBH}:
\begin{equation*}
\tilde{M} = \frac{M_{\gstar}}{3.83\times(\kovermb)^{\frac{2}{3}}},
\end{equation*}
where also here the mass threshold for QBH production, $M_{\mathrm{th}}$, is assumed to be equal to the higher-dimensional Planck scale.

Previous limits on the types of QBH production described in this article were set by the \mbox{ATLAS} experiment using final states with an energetic photon and a
jet \cite{ATLAS_qbh_gj} as well as final states with an energetic lepton and a jet \cite{Aad:2013gma}. 
Previous limits also exist from the CMS experiment from a search dominated by multi-jet final states~\cite{cmsDijet_qbh}.  The \mbox{ATLAS} experiment has also set limits on the production of a different type of QBHs using dijet events~\cite{blackmax,ATLAS_qbh_dijet}.  
While QBHs are not resonances, an increase in the dilepton production cross-section near the black hole threshold is expected.  
The expected signal is therefore similar to that predicted by resonance models, and QBHs are thus referred to as resonances in the remainder of this article.


\subsection{\label{sec:mwt_section}Minimal Walking Technicolor}
\label{introMWT}

Another solution to the hierarchy problem is to postulate that the Higgs boson is a composite particle, bound by a strong force called technicolor.  Technicolor models use the new strong dynamics to break electroweak symmetry.  These models predict the existence of new narrow technimeson resonances with masses of a few hundred GeV decaying to the dilepton final state. The interpretation used here is in the context of the Minimal Walking Technicolor model~\cite{TC3,TC4,TC5,Foadi:2012bb}, which predicts a composite Higgs boson having properties consistent, within current uncertainties, with the Higgs boson discovered at the LHC \cite{atlasHiggs,cmsHiggs}. 

The MWT model used here is defined by the following parameters: the bare axial-vector and vector masses, $M_A$ and $M_V$; the coupling of the spin-1 resonance to SM fermions $g/\gtilde$, where $g$ is the coupling constant of the weak interaction and \gtilde\ is the strength of the spin-1 resonance interaction; the $S$-parameter obtained using the zeroth Weinberg Sum Rule, used to constrain $M_A$ and $M_V$; the Higgs boson mass $m_H$, and $s$, the coupling of the Higgs boson to composite spin-1 states.  Here the $S$-parameter and $s$ are set according to the recommendation set forth in Ref.~\cite{Andersen:2011nk}: $S=0.3$ and $s=0$, while $m_H=125$~GeV is used for the Higgs boson mass.  The physical mass of about 125~GeV for the Higgs boson emerges after top quark corrections are taken into account \cite{Foadi:2012bb}.

This model predicts new particles in the form of technimeson triplets: $\Rone^{0,\pm}$ and $\Rtwo^{0,\pm}$.  The $\Rone^0$ and $\Rtwo^0$ are produced by quark--antiquark annihilation and decay to dilepton final states via an intermediate $Z/\gamma^*$ state.  For each pair of values ($M_{\Rone}$, \gtilde), the values of $M_{\Rtwo}$, $M_A$ and $M_V$ are unique.  The widths and the mass difference of \Rone\ and \Rtwo\ vary strongly depending on the model parameters~\cite{Belyaev:2008yj}.  In this analysis, the model parameter $\gtilde=2$ is used.  Previous studies have shown \cite{zprime_2011_paper} that the \mll\ distributions obtained with $\gtilde=2$ are representative of those for all values of \gtilde\ and $M_A$ to which this analysis is currently sensitive.  For this analysis, an \mll\ distribution accounting for contributions from both \Rone\ and \Rtwo\ is used.  However, the magnitude of the mass difference between the two and the characteristics of the distribution are dependent on \gtilde\ and $M_A$.  For larger values of \gtilde\ and small values of $M_A$, \Rtwo\ is broad with a reduced amplitude, and therefore does not contribute significantly to the signal shape.

Previous limits on this model were set by \mbox{ATLAS} on the bare axial mass, $M_A$, in the MWT model.
For a value of the coupling parameter $\gtilde=2$, $M_A$ values less than 1.57~TeV were excluded at 95\%~CL~\cite{zprime_2011_paper}.


\section{\label{sec:atlas}ATLAS detector}

The \mbox{ATLAS} detector~\cite{atlas:detector}  
consists of an inner tracking detector system (ID) surrounded by a superconducting solenoid, 
electromagnetic and hadronic calorimeters, and a muon spectrometer (MS). 
Charged particles in the pseudorapidity\footnote{ATLAS uses a right-handed coordinate system with its origin at the nominal 
interaction point in the center of the detector and the $z$-axis along the beam pipe. The $x$-axis points from the interaction point to the 
center of the LHC ring, and the $y$-axis points upward. Cylindrical coordinates $(r,\phi)$ are used in the transverse plane, 
$\phi$ being the azimuthal angle around the beam pipe. The pseudorapidity is defined in terms of the polar angle 
$\theta$ as $\eta=-\ln\tan(\theta/2)$.}
range $|\eta| < 2.5$ are reconstructed with the ID, which consists of layers of silicon pixel and microstrip 
detectors and a straw-tube transition-radiation tracker having coverage within $|\eta| < 2.0$.
The ID is immersed in a 2~T magnetic field provided by the solenoid.
The latter is surrounded by a hermetic calorimeter that covers $|\eta| < 4.9$
and provides 3-dimensional reconstruction of particle showers.
The electromagnetic calorimeter is a liquid argon sampling calorimeter, 
which uses lead absorbers for $|\eta| < 3.2$ and copper absorbers in the very forward region.
The hadronic sampling calorimeter uses plastic scintillator tiles as the active material and iron absorbers in the region $|\eta| < 1.7$.
In the region $1.5 < |\eta| < 4.9$, liquid argon is used as active material, with copper or/and tungsten absorbers.
Outside the calorimeter, air-core toroids supply the magnetic field for the MS. 
There, three stations of precision chambers allow the accurate measurement of muon track curvature in the region $|\eta| < 2.7$.
The majority of these precision chambers are composed of drift tubes,
while cathode strip chambers provide coverage in the inner stations of the forward region for $2.0 < |\eta| < 2.7$.
Additional muon chambers installed between the inner and middle stations of the forward region and commissioned prior to the 2012 run improve measurements in the transition region of $1.05<|\eta|<1.35$ where the outer stations have no coverage.
Muon triggering is possible in the range $|\eta| < 2.4$, using resistive-plate chambers in the central region and thin-gap chambers in the forward region.
A three-level trigger system~\cite{atlas_trigger} selects events to be recorded for offline analysis.


\section{\label{sec:data}Data sample}

The events in the dataset were collected 
during periods with stable beams and all relevant subsystems operational.
The $pp$ collision data recorded between April and December 2012 at $\sqrt{s}=8$ \tev\ 
amount to \LumiElectronsInFb\ in the dielectron channel and \LumiMuonsInFb\ in the dimuon channel.

In the dielectron channel, events are triggered by the presence of two 
energy deposits (``clusters") in the electromagnetic calorimeter,
one with transverse momentum ($\pt$) threshold of $\pt > 35$~\gev, and the other  with $\pt > 25$~GeV.
The shower profiles are required to be consistent with those expected for electromagnetic showers~\cite{atlas:egamma_perf2011}.
This trigger is preferred over a dedicated dielectron trigger, which incorporates tracking information, because it is advantageous in the estimation of the data-driven background, as explained in Sec.~\ref{sec:data_driven}.
In the dimuon channel, events are triggered by at least one of two single-muon triggers with transverse momentum thresholds of $\pt > 24$~\gev\ or $\pt > 36$~\gev\ with an additional requirement that the muon candidate be isolated (see Sec.~\ref{sec:leprec}) for the former case.


\section{\label{sec:sim}Simulated samples}

Expected signal and background yields, with the exception of 
certain data-driven background estimates,
are evaluated with simulated Monte Carlo (MC) samples and normalized 
using the highest-order 
cross-section predictions available in perturbation theory.

The sample used to model the Drell--Yan ($q\bar{q}\rightarrow Z/\gamma^*\rightarrow \ll$) background 
is generated at next-to-leading order (NLO) using \powheg~\cite{Alioli:2010xd} and the CT10 PDF~\cite{CT10}, 
with \pythia~8~\cite{pythia8} to model parton showering and hadronization. 
For this and all other samples, the final-state photon radiation (FSR) is handled by \photos~\cite{fsr_ref}, and the interaction of particles with the detector
and its response are modeled using a full \mbox{ATLAS} detector simulation~\cite{atlas:sim} based on \geant4~\cite{geant}.
The $Z/\gamma^*$ differential cross-section with respect to mass has been calculated at next-to-next-to-leading-order (NNLO) perturbative QCD (pQCD) 
using FEWZ~\cite{fewz,fewz1} with the MSTW2008NNLO PDF~\cite{mstw}.
The calculation includes NLO electroweak (EW) corrections beyond FSR, as well as 
a contribution
from the irreducible, non-resonant photon-induced (PI) background, $\gamma\gamma \to \ll$. 
The PI contribution is estimated using the MRST2004qed PDF~\cite{MRST2004QED} at leading order (LO),
by taking 
an average of the predictions obtained under the current and
constituent quark mass schemes. 
Differences between the average and the individual results from those schemes
are used to assign the uncertainty on this additive correction.
The PI corrections were verified by SANC~\cite{Bardin:2012jk,Bondarenko:2013nu}. 
An additional small correction arises from single boson production
in which the final-state charged lepton radiates a real $W$ or $Z$ boson.
This was estimated using \madgraph~\cite{madgraph5}, following the prescription
outlined in Ref.~\cite{ewrad}.
A mass-dependent $K$-factor used to scale the $Z/\gamma^{\ast}$ background
samples is obtained from the ratio of the calculated NNLO pQCD cross-section,
with additional EW, PI and real $W/Z$ corrections,
to the cross-section from the \powheg\ sample.
The values of the $K$-factors as evaluated at dilepton masses of 1, 2 and 3~TeV are 1.07, 1.10 and 1.14, respectively.

Other important backgrounds are  
due to diboson ($WW$, $WZ$ and $ZZ$) and top quark production.
The diboson processes are generated with \herwig~\cite{herwig,Corcella:2002jc}, using the CTEQ6L1 PDF~\cite{Pumplin:2002vw}. 
The diboson cross-sections are known to NLO with an uncertainty of 5\%, and the values 
used are 57~pb ($WW$), 21~pb ($WZ$) and 7.4~pb ($ZZ$), as calculated with MCFM~\cite{Campbell:1999mcfm}. 
Backgrounds from \ttbar\ and from single top production in association with a \w\ boson
are modeled with \mcatnlo~\cite{mcatnlo,Frixione:2003ei,Frixione:2008yi}
with \herwig\ using the CT10 PDF.
The $t\bar{t}$ cross-section 
is 
$\sigma_{t\bar{t}}= 253^{+13}_{-15}$~pb
for a top quark mass of $172.5 \gev$. This is calculated at NNLO in QCD 
including resummation of next-to-next-to-leading logarithmic soft gluon terms with 
{\sc Top}++2.0 \cite{Cacciari:2011,Baernreuther:2012ws,Czakon:2012zr,Czakon:2012pz,Czakon:2013goa,TopPP:2011}. 
The PDF and $\alpha_S$ uncertainties on the $t\bar{t}$ cross-section are calculated using the PDF4LHC prescription~\cite{Botje:2011sn}
with the MSTW2008 68\% CL NNLO \cite{mstw,mstw_alphas}, CT10 NNLO  \cite{CT10,Gao:2013xoa} and NNPDF2.3 5f FFN \cite{nnpdf23} PDF error sets 
added in quadrature to the scale uncertainty. Varying the top quark mass by $\pm$1~GeV leads to an additional systematic 
uncertainty of 
+8~pb and --7~pb, 
which is also added in quadrature.
The single top background in association with a $W$ boson has a cross-section of
$\sigma_{Wt}= 22.4 \pm 1.5$~pb \cite{Kidonakis:2010ux}.
Given that the $Wt$ contribution is small compared to the $t\bar{t}$ cross-section, an overall
uncertainty of 6\% is estimated 
on the top quark background.
The simulated top quark samples are statistically limited at
high invariant mass, and the expected number of events as a function of \mll\ is therefore extrapolated into this region using fits.
A number of fits to the invariant mass distribution are carried out, exploring various fit ranges as well as the two fit 
functions $y(x) = p_1 \, x^{p_2 + p_3 \log x}$ and $y(x) = p_1 / (x + p_2)^{p_3}$, where $y$ represents the expected yield and $x=\mll$.
The
mean and RMS of these fits are used as the background contribution and its uncertainty, respectively.
Background contributions from events with jets or photons in the final state that pass the electron selection criteria
are determined using the data,  
as explained in Sec.~\ref{sec:data_driven}.
In the muon channel this background is negligible.
In order to avoid double counting, the simulated samples in the electron channel are filtered for the presence of two electrons.

An overview of the simulated MC signal and background samples is given in Table~\ref{tab:mc}. 
\begin{table}[h]
\caption{Overview of simulated samples used.}
\label{tab:mc}
\centering
\small
\begin{tabular}{c|lll}
\hline
\hline
Process          			& Generator 				& Parton shower 	& PDF 		\\
\hline
Drell-Yan				& \powheg   				& \pythia~8.162		& CT10	   	\\
Diboson					& \herwig++~2.5.2			& \herwig~6.520  	& CTEQ6L1	\\
\ttbar, $Wt$				& \mcatnlo~4.06   			& \herwig~6.520	& CT10		\\
\zp					& \pythia~8.165  		& \pythia~8.165		& MSTW2008LO	\\
\gstar					& \pythia~8.160  		& \pythia~8.160		& CTEQ6L	\\
\zstar					& \calchep~4.5.1 		& \pythia~8.165 	& MSTW2008LO	\\
MWT					& \madgraph      & \pythia~8.165		& MSTW2008LO	\\
QBH					& QBH 1.05  				& \pythia~8.165 	& CT10		\\
\hline
\hline
\end{tabular}
\end{table}


Simulated signal processes for the \zp\ models are obtained by reweighting \pythia~8 Drell--Yan samples to the shape of the resonance.
The same technique is used for MWT signals, and the shape of the resonance is obtained using \madgraph.
A reweighting procedure is also used for \zstar\ and \gstar\ signals,
but it is applied to dedicated samples generated with \calchep~\cite{calchep} in the case of \zstar, and with \pythia~8 in the case of \gstar.
For the QBH signals, samples are generated for each assumed energy threshold ($M_{\mathrm{th}}$)
using the QBH~\cite{ref:Gingrich_generator} generator.
The MSTW2008LO PDF~\cite{mstw} is used for all signal samples, except the \gstar, which uses the CTEQ6L PDF~\cite{Pumplin:2002vw}.
The ratio of the NNLO pQCD cross-section calculated with FEWZ without the additional EW, PI and real $W/Z$ corrections
to the cross-section from the \pythia~8 sample 
is used to determine a mass-dependent $K$-factor for the signal samples.
The values of the $K$-factors as evaluated at dilepton masses of 1, 2 and 3~TeV are 1.22, 1.16 and 1.16, respectively.  
The additional EW and real $W/Z$ corrections are not applied to the signal samples because
the dominant EW corrections depend on the $W$ and $Z$ boson couplings
of the new particle, and are therefore model-dependent. The PI contribution is 
non-resonant and thus only contributes to the background.  
No $K$-factor is applied to the leading-order \zstar\ and QBH cross-sections.  This is due to the
different coupling of the \zstar\ to fermions, and the unknown gravitational interaction.
For \gstar, a NLO $K$-factor was provided by the authors of Refs~\cite{Mathews:2004xp,Mathews:2005bw,Kumar:2006id}, 
using CTEQ6L, which is the same PDF used in the simulation of the signal.


\section{Lepton reconstruction}
\label{sec:leprec}
Electron candidates are formed from clusters of cells reconstructed in the electromagnetic calorimeter with an associated well-reconstructed ID track.
The track and the cluster must satisfy a set of
identification criteria~\cite{atlas:egamma_perf2011} that are optimized for high 
pile-up\footnote{Multiple $pp$ collisions occurring in the same or neighboring bunch crossings.} conditions. 
These criteria require
the shower profiles to be consistent with those expected
for electrons and impose a minimum requirement on the 
amount of transition radiation. 
In addition, to suppress background from photon 
conversions, a hit in the first layer of the pixel detector is
required if an active pixel layer is traversed. 
The electron's energy is obtained from the calorimeter measurements and its direction from the associated track.

At transverse energies ($\et$) relevant to this search, the calorimeter energy resolution is measured in data to be 1.2\% for electrons in the central region ($|\eta| < 1.37$) and 1.8\% in the 
forward region ($1.52 < |\eta| \leq 2.47$)~\cite{atlas:egamma_perf}.
For dielectron masses above 200~GeV, the mass resolution is below 2\% over the entire $\eta$ range.

To suppress background from misidentified jets, isolated electrons are selected.
A limit is placed on the energy, corrected for transverse shower leakage and pile-up, contained 
in a cone of radius $\Delta R = 0.2$ surrounding the electron candidate in the ($\eta,\phi$) plane: $\Delta R = \sqrt{(\Delta\eta)^2 + (\Delta\phi)^2}$.  
For the leading (highest-$\pt$) electron candidate this energy is required to be less than $0.007\times \et + 5.0$~GeV, 
while for the subleading electron candidate a requirement of less than $0.022\times \et + 6.0$~GeV is used. 
These requirements have been optimized to maintain a high selection efficiency of $\approx 99$\% for each electron candidate.  The difference in the isolation selection for the leading and subleading electrons takes into account the different energy losses due to bremsstrahlung.

Muon tracks are first reconstructed~\cite{Aad:2014zya,muon_reco_paper} separately in the ID and in the MS. 
The two tracks are then matched and a combined fit is performed using ID and MS hits, 
taking into account the effects of multiple scattering and energy loss in the calorimeters. 
The momentum is taken from the combined fit.
Each muon is required to have a minimum number of hits in each of the ID components.
To obtain optimal momentum resolution, at least one selected muon is required
to have at least three hits in each of three stations
of the MS, or, for muons in the very forward region, at
least two hits in the cathode strip chambers and at least
three hits in the middle and outer MS stations. At least
one hit in each of two layers of the trigger chambers is
also required.
These muons are referred to as 3-station muons, and have \pt\ resolution at 1~TeV ranging from 19\% to 32\%, depending on $\eta$.
In the very forward region of the MS, the hit requirement in the inner station corresponds to at least two hits in the cathode strip chambers. 

In addition to 3-station muons, the best remaining muon candidates in the central region of the MS ($|\eta|<$ 1.05)
with at least five precision hits in each of the inner and outer stations are selected, and are referred to as 2-station muons.  
These 2-station muons are required to have at least one hit in one layer of the trigger chambers,
and they have slightly worse \pt\ resolution than the 3-station muons.

Residual misalignments of the muon detectors, which could cause a degradation of the momentum resolution, were studied with collision data in which the muons traversed overlapping sets of muon chambers.
The effects of these misalignments and the intrinsic position resolution are included in the simulation.
Muon candidates passing through chambers where the alignment quality does not allow a reliable momentum measurement at high \pt\ are rejected.

For each 3-station (2-station) muon, the difference between the standalone momentum measurements from the ID and MS 
must not exceed five (three) times the sum in quadrature of the standalone uncertainties.
To suppress background from cosmic rays, the muons are also required to satisfy requirements on the track impact parameters
with respect to the primary vertex of the event.
The impact parameter along the beam axis is required to be within 1~mm, and the transverse impact parameter is required to be within 0.2~mm.
The primary vertex of the event is defined as the 
reconstructed vertex consistent with the beam spot position with the highest $\sumpt^2$. 
The sum includes the $\pt^2$ of all tracks associated with the primary vertex.
At least three associated tracks are required, each with \pt\ above 0.4~GeV.
To reduce the background from misidentified jets, each muon is required to be isolated such that $\Sigma\pt(\Delta R < 0.3)/\pt(\mu)<0.05$, 
where $\Sigma \pt (\Delta R < 0.3)$ is the scalar sum of the \pt\ of all other tracks with $\pt >$ 1 GeV within a cone of radius $\Delta R = 0.3$ around the direction of the muon.


\section{Event selection}

Events are required to have at least one reconstructed primary vertex.

For the dielectron channel, at least two reconstructed electron candidates within $|\eta| < 2.47$ are required. 
The leading and subleading electron must satisfy $\et > 40$~\gev\ and $\et > 30$~\gev, respectively. 
The transition region between the central and forward regions of the calorimeters, in the range $1.37 \leq |\eta| \leq 1.52$, exhibits degraded energy resolution and is therefore excluded. 
Because of possible charge misidentification, an opposite-charge requirement is not placed on electron candidates.  Charge misidentification can occur either due to bremsstrahlung, or due to the limited momentum resolution of 
the ID at very high \pt. 

\begin{figure}[t]
  \centering
\includegraphics[width= \columnwidth]{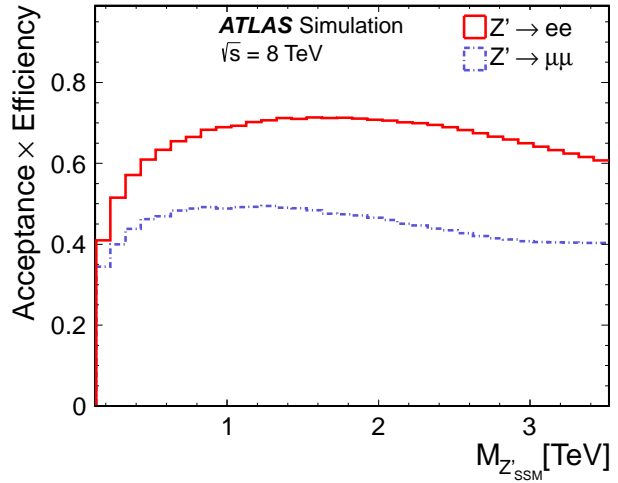} 
  \caption{Product of acceptance and efficiency for the dielectron (upper distribution) and dimuon (lower distribution) selections as a function of the \zpssm\ pole mass. 
  }
  \label{fig:SigEff_truem}
\end{figure}

The product of acceptance and efficiency ($A \times \epsilon$) is defined as the fraction of simulated candidate events that pass the dilepton event selection 
requirement in the \mll\ search region 128~GeV $< \mll <$ 4500~GeV, out of those generated with a Born level dilepton mass greater than 60~GeV. 
Fig.~\ref{fig:SigEff_truem} shows $A \times \epsilon$ as a function of the \zpssm\ pole mass for both channels. 
Using the described search criteria, $A \times \epsilon$ in the dielectron channel is found to be 71\% for a \zpssm\ pole mass of 2~\tev.
For low values of the \zpssm\ pole mass, $A \times \epsilon$ rises due to kinematic selection requirements. 
It drops again at high pole mass because the strong decrease of the parton luminosity at high momentum transfer 
enhances the relative fraction of events in the low-mass tail of the spectrum arising from off-shell \zpssm\ production.

Muons passing the reconstruction criteria are required to satisfy $\pt > 25$~GeV
and are used to build opposite-charge muon pairs.  
If two opposite-charge muons passing the 3-station selection are found,
they are used to make the pair and the event is said to pass the ``primary dimuon selection.''
If no primary dimuon candidate is found, pairs are built with one 3-station muon and a 2-station muon of opposite charge.
Events with such pairs are said to pass the ``secondary dimuon selection.''  
For both selections, if more than one dimuon candidate is found in an event, 
the one with the highest transverse momentum scalar sum is selected.
In the case of a \zpssm\ of mass 2~\tev, $A \times \epsilon$ in the dimuon channel is estimated to be 46\%, as 
can be seen in Fig.~\ref{fig:SigEff_truem}. The contribution of the primary (secondary) dimuon selection is about 42\% (4\%) at 2~\tev.
Due to the stringent requirements placed on the number and distribution of hits required in the MS, which ensure good momentum resolution at large~\mll, the $A \times \epsilon$ for the dimuon channel is lower compared to the dielectron channel.

\section{Data-driven backgrounds}
\label{sec:data_driven}

As mentioned above, background contributions from events with jets or photons in the final state that pass 
the electron selection criteria 
are determined using the data.
This includes contributions from dijet, heavy-flavor quarks and $\gamma$ + jet production, 
referred to hereafter as the dijet background. Additional contributions are due to \wpjet\ processes and 
top quark production with \wpjet\ final states, referred to hereafter as \wpjet\ background.

The probability that a jet is misidentified as an electron (the ``fake rate'') is determined as a function of \et\ and $\eta$ using background-enriched data samples.
These samples are recorded using several inclusive jet triggers with \et\ thresholds in the range 25--360~GeV.
In each of these samples, the fake rate $f_1$ ($f_2$) is calculated as the fraction
of leading (subleading) electron candidates that pass the nominal electron identification and
isolation requirements (``tight"), with respect to the entire sample of ``loose'' electron
candidates. 
The loose candidates satisfy only a
subset of the nominal electron identification criteria, which has to be
stricter than the trigger requirements imposed on a single object. 
To avoid bias due to a real electron contribution
from $W$ decays or the Drell--Yan process, events are vetoed in the following cases: if the
missing transverse momentum is larger than 25~GeV, if they contain two
identified electrons satisfying strict criteria or if they contain two electrons
satisfying less strict criteria but with an invariant mass between 71~GeV and 111~GeV.
A weighted average of the fake rates obtained from the jet samples
is then calculated. The values of the fake rates are around 10\%.  They are not strongly \et-dependent,
but are smaller at central pseudorapidities and increase to as high as 20\% for $2.4 < |\eta| < 2.47$.

In addition to the fake rate, the probability $r_1$ ($r_2$) that a real electron in the sample
of loose electrons satisfies the nominal electron identification and leading (subleading)
isolation requirements is used in evaluating this background.  This probability is computed from MC simulation.
Potential differences between data and simulated samples in lepton identification and isolation efficiencies are accounted for by 
applying scale factors to the simulation, which are generally close to unity.
The values for $r_1$ and $r_2$ are well above 90\% for all \et\ and $\eta$.

A system of equations is used to solve for the unknown true
contribution to the background
from events with one or more fake electrons.
The relation between the number of true paired objects $N_{ab}$, 
with $E^a_{\rm T} > E^b_{\rm T}$ and $a,b \in \{R,F\}$,
and the number of measured pairs in the triggered sample $N_{xy}$, with $x,y \in
\{T,L\}$, can be written as:
\begin{widetext}
\begin{equation}
\label{ff_matrix_full}
\begin{pmatrix} \NTT \\ \NTL \\ \NLT \\ \NLL \end{pmatrix}
=
  \begin{pmatrix}
  r_1 r_2 & r_1 f_2 & f_1 r_2 & f_1 f_2 \\
  r_1 (1-r_2) & r_1 (1-f_2) & f_1(1-r_2) & f_1 (1-f_2) \\
  (1-r_1)r_2 & (1-r_1)f_2 & (1-f_1) r_2 & (1-f_1)f_2 \\
  (1-r_1)(1-r_2) & (1-r_1)(1-f_2) & (1-f_1)(1-r_2) & (1-f_1)(1-f_2)
  \end{pmatrix}
  \begin{pmatrix} \NRR \\ \NRF \\ \NFR \\ \NFF \end{pmatrix}.
\end{equation} 
\end{widetext}

The subscripts $R$ and $F$ refer to real electrons and fakes (jets),
respectively. The subscript $T$ refers to electrons that pass the tight
selection. The subscript $L$ corresponds to
electrons that pass the loose requirements described above but
fail the tight requirements.

The background is given as the part of $\NTT$, the number of pairs
where both objects are reconstructed as signal-like, originating from a pair of objects with at least one fake: 
\begin{equation}
\label{bkg_true_quantities}
N_{TT}^{\rm{Dijet} \& \wpjet\ }  = r_1 f_2 N_{RF} + f_1 r_2 N_{FR} + f_1 f_2 N_{FF}.
\end{equation}
The true paired objects on the right-hand side of Eq.~(\ref{bkg_true_quantities}) can be expressed in terms of measurable 
quantities (\NTT, \NTL, \NLT, \NLL) by inverting the matrix in Eq.~(\ref{ff_matrix_full}).

The dijet background in the dimuon sample is evaluated from data by 
reversing the requirement that muons pass the track isolation requirement based on the 
variable $\Sigma \pt (\Delta R < 0.3) / \pt$.  The method is further described in Ref.~\cite{Aad:2011eps}. 
The contribution of the dijet background
in the dimuon channel is negligible, as is the background from cosmic rays.


\section{\label{sec:sys}Systematic uncertainties}

The treatment of systematic uncertainties in this analysis is
simplified by the fact that 
the backgrounds are normalized to the data in the region of the \z\ peak. 
This procedure makes the analysis insensitive to the uncertainty on
the measurement of the integrated luminosity as well as other mass-independent systematic uncertainties. 
A mass-independent systematic error of 4\% is assigned to the signal expectation due to the uncertainty on the $Z/\gamma^{\ast}$ cross-section in the normalization region.
This uncertainty is due to the PDF and \alphas\ uncertainties obtained from the 90\% CL MSTW2008NNLO PDF error set, 
using the program \vrap~\cite{vrap} in order to calculate the NNLO Drell--Yan cross-section in the normalization region.
In addition, scale uncertainties are estimated by varying the renormalization and factorization scales simultaneously 
up and down by a factor of two, also using VRAP.

Mass-dependent systematic uncertainties include theoretical and
experimental effects on the signal and background.
These uncertainties are correlated across all \mll\ bins in the search region. 
The mass-dependent theoretical uncertainties are applied to the $Z/\gamma^{\ast}$ background expectation only.
In general, theoretical uncertainties are not applied to the signal. However, the mass dependence of the PDF uncertainty due to acceptance variations was checked and found to be negligible.
It is assumed that the experimental uncertainties are fully correlated between the signal and all types of background.
In the statistical analysis, all systematic uncertainties estimated to have an impact
$< 3\%$ on the expected number of events for all values of \mll\ are neglected, as they have negligible impact on the results of the search.

The combined uncertainty on the $Z/\gamma^{\ast}$ background due to PDF (``PDF variation'') and \alphas\ is obtained from the 90\% CL MSTW2008NNLO PDF error set, 
using \vrap\ in order to calculate the NNLO Drell--Yan cross-section as a function of \mll. 
The resulting uncertainties at dilepton masses of 2~TeV and 3~TeV are given
in Tables~\ref{tab:systematicSummary_2TeV} and \ref{tab:systematicSummary_3TeV}, respectively.
An additional uncertainty is assigned to take into account potential differences 
between modern PDFs
at the same \alphas = 0.117:
MSTW2008NNLO, CT10NNLO, NNPDF2.3~\cite{nnpdf23}, ABM11~\cite{Alekhin:2012ig} and HERAPDF1.5~\cite{herapdf10}.
Of these, only the central values for ABM11 fall outside of the MSTW2008NNLO PDF's uncertainty band.
Thus, an envelope of the latter uncertainty and the ABM11 central value is formed with respect to the central value of the MSTW PDF.
The 90\% CL uncertainty from MSTW is subtracted in quadrature from this envelope, 
and the remaining part, which is only non-zero when the ABM11 central value is outside the MSTW2008NNLO PDF uncertainty,
is quoted as ``PDF choice.''
Scale uncertainties are estimated by varying the renormalization and factorization scales simultaneously 
up and down by a factor of two, also using VRAP. The resulting maximum variations are taken as uncertainties
and are less than 3\%.
The uncertainty on the PI correction is taken as half the difference 
between the predictions obtained under the current and constituent quark mass schemes, 
as discussed in Sec.~\ref{sec:sim}.
In addition, a systematic uncertainty is attributed to EW corrections for both channels,
corresponding to the difference in the theoretical calculation between
FEWZ and SANC.

On the experimental side, a systematic effect common to both channels is due to an uncertainty of 0.65\% on the
beam energy \cite{beam_energy}. The effect on the background cross-section was evaluated for the dominant $Z/\gamma^{\ast}$ 
background only, and it can be as high as 5\% at high dilepton masses.  
For the signals considered here, the effect of this uncertainty on $A \times \epsilon$ is negligible ($<$1\%).

In the dielectron channel,  the systematic uncertainty is dominated by the determination of  background  contributions 
with jets faking electrons in the final state, mainly dijet and \wpjet\ processes.
In order to derive this uncertainty, the method described above was altered by assuming $r_1 = r_2 = 1$. 
This second ``matrix method'' leads to a simplification of the matrix in Eq.~(\ref{ff_matrix_full}), 
but also necessitates the use of MC corrections for the identification and isolation inefficiencies of real electrons.
Large corrections from MC simulation can be avoided in a third ``matrix method'' where 
objects in the background-enriched sample fail the requirement on the matching
between track and cluster, instead of the full identification and
isolation requirements.

In addition to the standard background-enriched sample recorded using the jet triggers, 
two alternative background-enriched samples are obtained using a ``Tag and Probe'' technique
on the jet-triggered sample and the sample triggered by electromagnetic objects. Here the choice of an electromagnetic-object trigger
that is looser than a dedicated electron trigger (see Sec.~\ref{sec:data}) leads to an enlarged 
sample.
The background-enriched sample of probes is obtained
by selecting a jet-like tag and a probe with 
the same charge, among other requirements, in order to
suppress real electron contamination. 
Finally, the default method and the two additional matrix methods are each used in conjunction with the default sample and 
the two different background-enriched samples, leading to nine different background estimates.
In the \mll\ search region, the maximum deviation of the eight alternative estimates from the default background estimate 
is 18\% and is taken as a systematic uncertainty at all values of \mll.

Furthermore, the different requirements used to suppress real electron contamination in the default fake-rate calculation are varied.
The largest deviations, about 5\%, occur when the value of the missing energy requirement is changed.
The statistical uncertainty on the fake rates results in an uncertainty on the background of at most 5\%.

Another systematic uncertainty can arise if fake rates are different for the various processes contributing to the background, and if the relative contributions of these processes in the data samples from which the fake rates are measured and in the data sample to which the fake rates are applied are different. 
Jets originating from bottom quarks have a higher fake rate than jets originating from light-quark jets, but the effect of this is negligible as the number of $b$-jets is small and similar in both samples.
As an additional check, the background is recalculated using all nine methods discussed above, but with separate fake rates for different background processes. 
The mean of these nine methods is in agreement with the background estimate from the default method.

Thus, adding the different sources of uncertainty in quadrature, an overall systematic uncertainty of 20\% is
assigned to the dijet and \wpjet\ background. 
At low invariant masses there is an additional uncertainty due to the statistical uncertainty from the sample to which the fake rates are applied. 
At high invariant masses this component is replaced by a systematic uncertainty due to the background extrapolation into this region.
The extrapolation is done in the same way as for the top quark background (see Sec.~\ref{sec:sim}) and dominates the uncertainty on 
the dijet and \wpjet\ background contribution at the highest invariant masses.

Experimental systematic uncertainties from
the electron reconstruction and identification efficiencies,
as well as from the energy calibration 
and resolution are neglected, as they alter the expected number 
of events by less than 3\%.

For the dimuon channel, the combined uncertainty on the trigger and reconstruction efficiencies is negligible.  
Inefficiencies may occur for muons with large energy loss due to bremsstrahlung in the outer parts of the calorimeter, 
interfering with muon reconstruction in the MS.
However, such events are rare and the corresponding systematic uncertainty is negligible over the entire mass range considered.
This is an improvement on previous \mbox{ATLAS} publications \cite{zprime_2011_paper}, which used a very conservative, and much larger, estimate: 6\% at 2~TeV.
In addition, the uncertainty on the resolution due to residual misalignments in the MS propagates 
to a change in the steeply falling background shape at high dilepton mass and in the width of signal line shape. 
The potential impact of this uncertainty on the background estimate reaches 3\% at 2~TeV and 8\% at 3~TeV. 
The effect on the signal is negligible.
As for the dielectron channel, the momentum scale uncertainty has negligible impact in the dimuon channel search.

Mass-dependent systematic uncertainties that change the expected number of events by at least 3\% anywhere in the $\mll$ distribution
are summarized 
in Tables~\ref{tab:systematicSummary_2TeV} and \ref{tab:systematicSummary_3TeV} for 
dilepton invariant masses of 2~TeV and 3~TeV, respectively. 
\begin{table}
\caption{
Summary of systematic uncertainties on the expected numbers of events at a dilepton mass of $\mll=2$~TeV, where
\NA\ indicates that the uncertainty is not applicable.
Uncertainties $<$\ 3\% for all values of \mee\ or \mmm\ are 
neglected in the respective statistical analysis.
}
\label{tab:systematicSummary_2TeV}
\centering
\addtolength{\tabcolsep}{+1pt}
\small
\begin{tabular}{l|cc|cc}
\hline
\hline
Source ($\mll=2$~TeV)      		& \multicolumn{2}{c|}{Dielectrons}    & \multicolumn{2}{c}{Dimuons} \\
                	    		& Signal & Backgr.                                 &  Signal & Backgr. \\
\hline
Normalization				& 4\%     & \NA			& 4\%   & \NA				\\
PDF variation				& \NA      & 11\%		& \NA    & 12\%	 			\\
PDF choice				& \NA      & 7\%			& \NA    & 6\%	 			\\
$\alpha_{s}$				& \NA      & 3\%			& \NA    & 3\%				\\
Electroweak corr.			& \NA      & 2\%			& \NA    & 3\% 				\\
Photon-induced corr.		& \NA      & 3\%			& \NA    & 3\% 				\\
Beam energy                            	& $<$\ 1\%      & 3\%                 & $<$\ 1\%  	& 3\%			 	\\  
Resolution				& $<$\ 3\%          & $<$\ 3\%                & $<$\ 3\%     & 3\%			\\
Dijet and \wpjet\   			& \NA      & 5\%			& \NA    & \NA 				\\
\hline		       		         				      			 
 Total					&  4\%    & 15\%		& 4\%   & 15\%			\\
\hline
\hline
\end{tabular}
\end{table}

\begin{table}
\caption{
Summary of systematic uncertainties on the expected numbers of events at a dilepton mass of $\mll=3$~TeV, where
\NA\ indicates that the uncertainty is not applicable. 
Uncertainties $<$\ 3\% for all values of \mee\ or \mmm\ are 
neglected in the respective statistical analysis.
}
\label{tab:systematicSummary_3TeV}
\centering
\addtolength{\tabcolsep}{+1pt}
\small
\begin{tabular}{l|cc|cc}
\hline
\hline
Source ($\mll=3$~TeV)          		& \multicolumn{2}{c|}{Dielectrons}    & \multicolumn{2}{c}{Dimuons} \\
                	    		& Signal  & Backgr.        &  Signal & Backgr. \\
\hline
Normalization				& 4\%     & \NA			& 4\%   & \NA				\\
PDF variation				& \NA      & 30\%		& \NA    & 17\%	 			\\
PDF choice				& \NA      & 22\%		& \NA    & 12\%	 			\\
$\alpha_{s}$				& \NA      & 5\%			& \NA    & 4\%				\\
Electroweak corr.			& \NA      & 4\%			& \NA    & 3\% 				\\
Photon-induced corr.		& \NA      & 6\%			& \NA    & 4\% 				\\
Beam energy                            	& $<$\ 1\%      & 5\%                	& $<$\ 1\%  	& 3\%   			\\  
Resolution				& $<$\ 3\%       & $<$\ 3\%                	& $<$\ 3\%  & 8\%		\\
Dijet and \wpjet\  			& \NA      & 21\%		& \NA    & \NA 				\\
\hline		       		         				      			 
 Total					&  4\%    & 44\%		& 4\%   & 23\% 			\\
\hline
\hline
\end{tabular}

\end{table}


\section{\label{sec:comp}Comparison of data and background expectations}

The observed invariant mass distributions, \mee\ and \mmm, are compared to the expectation from SM backgrounds
after final selection. 
To make this comparison, the sum of all simulated backgrounds, with the relative contributions fixed
according to the respective cross-sections, is scaled such that the result
agrees with the observed number of data events in the 80 - 110 GeV normalization region, 
after subtracting the data-driven background in the case of the electron channel.
The scale factors obtained with this procedure are 1.02 in the dielectron channel and 0.98 in the dimuon channel.
It is this normalization approach that allows the mass-independent uncertainties to cancel in the statistical analysis.

\begin{figure}
  \centering
  \includegraphics[width=\columnwidth]{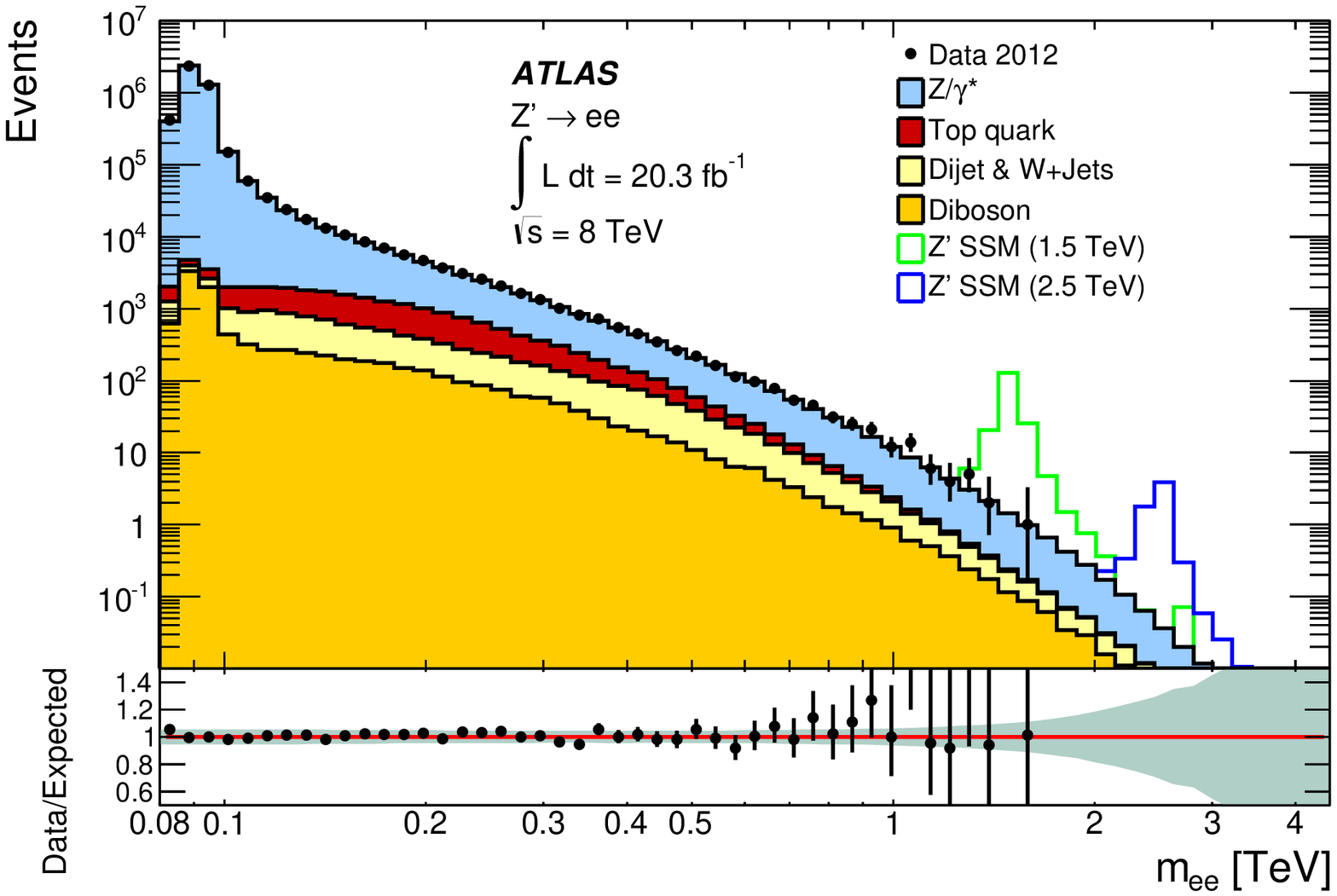}
  \includegraphics[width=\columnwidth]{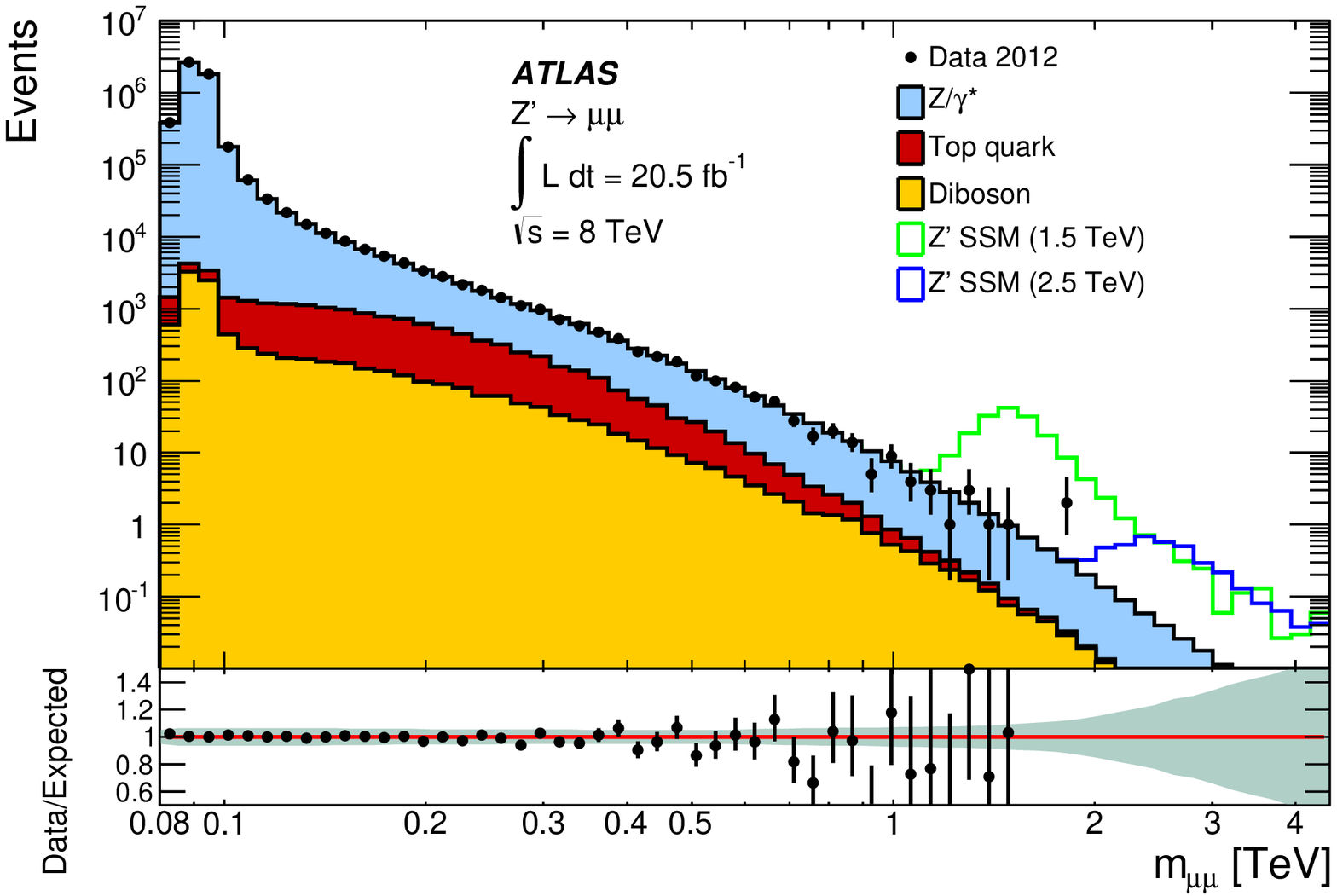}
  \caption{ Dielectron (top) and dimuon (bottom) invariant mass (\mll ) distributions after event selection, 
  with two selected \zpssm\ signals overlaid, compared to the stacked sum of all expected backgrounds,
  and the ratios of data to background expectation. The bin width is constant in $\log \mll$. The green band in the ratio plot 
  shows the systematic uncertainties described in Sec.~\ref{sec:sys}. }
  \label{fig:mll_ssm}
\end{figure}

Figure~\ref{fig:mll_ssm} depicts the \mll\ distributions for the dielectron and dimuon final states. 
The bin width of the histograms is constant in $\log \mll$, chosen such that a possible signal peak spans multiple bins 
and the shape is not impacted by statistical fluctuations at high mass.
The shaded band in the ratio inset represents the systematic uncertainties described in Sec.~\ref{sec:sys}. 
Figure~\ref{fig:mll_ssm} also displays the expected \zpssm\ signal for two mass hypotheses. 
Table~\ref{tab:backgroundTable} 
shows the number of data events and the estimated backgrounds in several bins of reconstructed dielectron and 
dimuon invariant mass above 110~GeV.
\begin{table*}[t!]
\fontsize{3mm}{3.5mm}\selectfont
\setlength{\abovecaptionskip}{6pt}
\setlength{\belowcaptionskip}{6pt}
\caption{The numbers of expected and observed events in the dielectron (top) and dimuon (bottom) channel in bins of 
the invariant mass \mll. The region 80--110~GeV is used to normalize the total background to the data. 
The errors quoted are the combined statistical and systematic uncertainties.}
\label{tab:backgroundTable}
\begin{tabular}{lcccccc}
\hline
\hline
$m_{ee}$ [GeV] & 110--200 & 200--400 & 400--800 & 800--1200 & 1200--3000 & 3000--4500\\
\hline
$Z/\gamma^{\ast}$  	& $ 122000 \pm 7000 $  & $ 14000 \pm 800 $  & $ 1320 \pm 70 $  & $ 70 \pm 5 $  & $ 10.0 \pm 1.0 $  & $ 0.008 \pm 0.004 $ \\
Top 			& $ 8200 \pm 700 $  & $ 2900 \pm 500 $  & $ 200 \pm 80 $  & $ 3.1 \pm 0.8 $  & $ 0.16 \pm 0.08 $  & $ < 0.001 $ \\
Diboson 		& $ 1880 \pm 90 $  & $ 680 \pm 40 $  & $ 94 \pm 5 $  & $ 5.9 \pm 0.4 $  & $ 1.03 \pm 0.06 $  & $ < 0.001 $ \\
Dijet \& $W$+jet 		& $ 3900 \pm 800 $  & $ 1290 \pm 320 $  & $ 230 \pm 70 $  & $ 9.0 \pm 2.3 $  & $ 0.9 \pm 0.5 $  & $ 0.002 \pm 0.004 $ \\
  \hline
Total 			& $ 136000 \pm 7000 $  & $ 18800 \pm 1000 $  & $ 1850 \pm 120 $  & $ 88 \pm 5 $  & $ 12.1 \pm 1.1 $  & $ 0.011 \pm 0.005 $ \\
  \hline
Observed		& $136200$ & $ 18986$ & $ 1862$ & $ 99$ & $ 9$ & $ 0$\\
\hline
\hline
   &    &    &    &  &  &  \\
\hline
\hline
$m_{\mu\mu}\mathrm{[GeV]}$ & 110--200 & 200--400 & 400--800 & 800--1200 & 1200--3000 & 3000--4500 \\ 
\hline
$Z/\gamma^{\ast}$  &	$ 111000 \pm 8000 $  &	$ 11000 \pm 1000 $  &	$ 1000 \pm 100 $  &	$ 49 \pm 5 $  &	$ 7.3 \pm 1.1 $  &	$ 0.034 \pm 0.022 $  \\ 
Top  &	$ 7100 \pm 600 $  &	$ 2300 \pm 400 $  &	$ 160 \pm 80 $  &	$ 3.0 \pm 1.7 $  &	$ 0.17 \pm 0.15 $  &	$ < 0.001 $  \\ 
Diboson  &	$ 1530 \pm 180 $  &	$ 520 \pm 130 $  &	$ 64 \pm 16 $  &	$ 4.2 \pm 2.1 $  &	$ 0.69 \pm 0.30 $  &	$ 0.0024 \pm 0.0019 $  \\ 
  \hline
Total  &	$ 120000 \pm 8000 $  &	$ 13700 \pm 1100 $  &	$ 1180 \pm 130 $  &	$ 56 \pm 6 $  &	$ 8.2 \pm 1.2 $  &	$ 0.036 \pm 0.023 $  \\ 
  \hline
Observed  &	$ 120011 $  &	$ 13479 $  &	$ 1122 $  &	$ 49 $  &	$ 8 $  &	$ 0 $  \\ 
\hline
\hline
\end{tabular}
\end{table*}
The number of observed events in the normalization region is 4,257,744 in the dielectron channel and 5,075,739 in 
the dimuon channel. The higher yield in the normalization region for the dimuon channel, despite the lower $A \times \epsilon$ at higher 
masses as displayed in Fig.~\ref{fig:SigEff_truem}, is due to the higher $E_{T}$ cuts on the electrons.  This reduces 
the yield in the dielectron channel in the region of the $Z$ peak.
The dilepton invariant mass distributions are well described by the predictions from SM processes.


\section{\label{sec:stats}Statistical analysis}

The data are compared to the background expectation in the search region. 
The comparison is performed by means of signal and background templates~\cite{Aad:2011xp, CDF:Zpmumu2fb} that 
provide the expected yield of events ($\bar{n}$) in each \mll\ bin. 
The dependence of the resonance width on the coupling strength is taken into account in the signal templates.
The coupling to hypothetical right-handed neutrinos and to $W$~boson pairs is neglected in the \zp\ search.
Interference of the \zp\ signal with the Drell--Yan background is taken into account in the Minimal \zp\ Models interpretation framework only.
When interference is not taken into account, $\bar{n}$ is given by $\bar{n} = n_X(\lambda, {\pmb\nu}) + n_{\dy} ({\pmb\nu}) + \nobg ({\pmb\nu})$,
where $n_{X}$ represents the number of events produced by the decay of a new resonance, $X$ 
($X=\zp, \zstar, \gstar, M_{\mathrm{th}}, R_{1,2}$), and 
$n_{\dy}$ and \nobg\ are the number of \dy\ (Drell--Yan) and other backgrounds events, respectively.
The symbol $\lambda$ represents the parameter of interest in the model, 
and ${\pmb\nu}$ is the set of Gaussian-distributed nuisance parameters incorporating systematic uncertainties.
When interference effects are included, $\bar{n} = n_{X+\dy}(\lambda, {\pmb\nu}) + \nobg ({\pmb\nu})$,
where $n_{X+\dy}$ is the number of signal plus \zgstar\ events and $X$ is the \zp\ boson in the Minimal Models interpretation.
A binned likelihood function is employed for the statistical analysis. The likelihood function is defined as the product 
of the Poisson probabilities over all mass bins in the search region,
\begin{equation*}
\mathcal{L(\lambda, {\pmb\nu})} = \prod_i^{N_{\rm{bins}}}\frac{e^{-\bar{n}_i}\bar{n}^{d_i}}{{d_i}!}G({\pmb\nu}).
\end{equation*}
The symbol $d_i$ corresponds to the observed number of events in bin $i$ of the \mll\ distribution and $G({\pmb\nu})$ represents 
the Gaussian functions for the set of nuisance parameters~${\pmb\nu}$.

The significance of a signal is summarized by a \pval, the probability of observing
an excess at least as signal-like as the one observed in data, assuming the null hypothesis.
The outcome of the search is ranked using a log-likelihood ratio (LLR) test statistic, using a \zpssm\ template assuming no interference.
Explicitly,
\begin{equation*}
\label{LLR_quation}
{\rm LLR} = -2\ {\rm ln}\ \frac{\mathcal{L} ({\rm data}\ |\ \hat{n}_{\zp}, \hat{M}_{\zp}, \hat{\pmb\nu} ) }{\mathcal{L} ({\rm data}\ |\ (\hat{n}_{\zp} = 0), \hat{\hat{\pmb\nu}} ) },
\end{equation*}
where $\hat{n}_{\zp}$ and $\hat{M}_{\zp}$ are the best-fit values for the \zp~normalization and the \zp~mass. 
The nuisance parameters that maximize the likelihood~$\mathcal{L}$ given the data are represented by $\hat{\pmb\nu}$ and $\hat{\hat{\pmb\nu}}$,
assuming in the numerator that a \zp\ signal is present, and in the denominator that no signal is present.
The LLR is scanned as a function of \zp\ cross-section and ${M}_{\zp}$ over the full considered mass range. This approach naturally includes the trials factor,
which accounts for the probability of observing an excess anywhere in the search region. 
The observed $p$-values for the dielectron and dimuon samples are \PvalueElectron \% and \PvalueMuon \%, respectively.

In the absence of a signal, upper limits on the number of events produced by the decay of a new resonance are determined at 95\%~CL. 
The same Bayesian approach~\cite{bayesianMethod} is used in all cases,
with a uniform positive prior probability distribution for the parameter of interest. When interference is not taken into account, the
parameter of interest is the signal cross-section times branching fraction (\xbr).
When interference effects are included the coupling strength is chosen as the parameter of interest, with a prior that is flat in the coupling strength to the fourth power.
The most likely number of signal events, and the corresponding confidence intervals, 
are determined from the binned likelihood function defined above. 
The product of acceptance and efficiency for the signal as a function of mass is different for each model 
considered due to different angular distributions, boosts, and line shapes.  
This is propagated into the expectation.
The dependence of the likelihood on the nuisance parameters is integrated out using the Markov Chain Monte Carlo method~\cite{bayesianMethod}.

The expected exclusion limits are determined using simulated pseudo-experiments with only SM processes 
by evaluating the 95\%~CL upper limits for each pseudo-experiment for each fixed value of the resonance pole mass, $M_{X}$.
The median of the distribution of limits is chosen to represent the expected limit. The ensemble of limits
is also used to find the 68\% and 95\% envelopes of the expected limits as a function of $M_{X}$.

The combination of the dielectron and dimuon channels is performed under the 
assumption of lepton universality. 
For each source of systematic uncertainty, the correlations across bins, as well as the correlations between signal and background, are taken into account.

\section{Model interpretation and results}
\label{results}

As no evidence for a signal is observed, limits are set in the context of the physics models introduced in Sec.~\ref{sec:ModelsSection}.  
For all but the Minimal \zp\ Models, limits are set on \xbr\ versus the resonance mass.
The predicted \xbr\ is used to derive limits on the resonance mass for each model.
Table~\ref{tab:sigma_br_vals} lists the predicted \xbr\ values for a few resonance masses and model parameters.
In the case of the \MM, limits are set on the effective couplings as a function of the resonance mass to incorporate interference effects 
of the \zp\ signal with the Drell--Yan background.
\begin{table}[bh]
\caption{
Values of \xbr\ for the different models. The model parameter $M$ corresponds to the mass of the \zp, \zpchi, \zppsi,
\zstar\ and \gstar\ boson. For the QBH models, $M=M_{\mathrm{th}}$ corresponds to the threshold mass, while for the MWT model $M=M_{\Rone}$.  The value $M=3$~TeV is not applicable for the MWT model, as the range of the limits is up to 2.25~TeV.}
\label{tab:sigma_br_vals}
\begin{center}
\begin{tabular}{lccc}
\hline
\hline
            				&  \multicolumn{3}{c}{\xbr\ [fb]}			\\
\hline			 	 	 	 	  	 
Model					& $M=1$ TeV 	& $M=2$ TeV 	& $M=3$ TeV 		\\
\hline			 	 	 	 	  	 
\zpssm					&170 		&3.4			&0.21		\\
$\zp_{\chi}$				&93		&1.5			&0.062		\\
$\zp_{\psi}$				&47		&0.87			&0.032		\\
\zstar					&300		&4.0			&0.076		\\
\gstar, \kovermb=0.1			&190		&1.8			&0.044		\\
RS QBH					&56		&0.40			&0.0065		\\
ADD QBH					&11000		&96			&1.8		\\
MWT, $\tilde{g}=2$			&31		&0.17			&\NA		\\
\hline
\hline
\end{tabular}
\end{center}
\end{table}


\subsection{Limits on narrow spin-1 \zpssm, $E_6$ \zp\ and \zstar\ bosons}

For the \zpssm, \esix-motivated \zp\ and \zstar\ bosons, the model specifies the boson's
coupling strength to SM fermions and therefore the intrinsic width. 
The parameter of interest in the likelihood analysis is therefore \xbr\ as a function of the new boson's mass.

\begin{figure}
  \centering
  \includegraphics[width=\columnwidth]{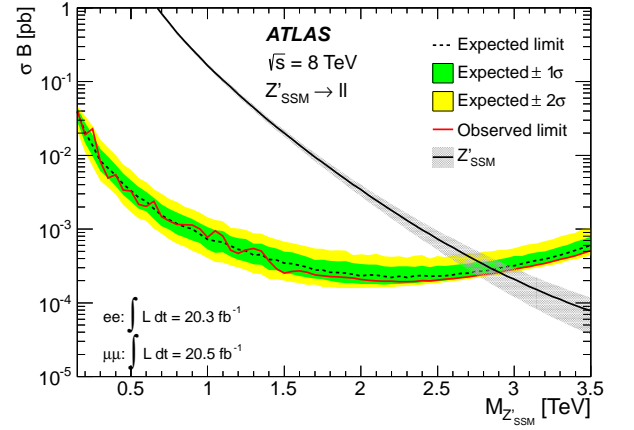}
  \caption{Median expected (dashed line) and observed (solid red line) 95\%~CL upper limits on cross-section times branching ratio (\xbr) in the combined dilepton channel, along with predicted \xbr\ for \zpssm\ production. The inner and outer bands show the range in which the limit is expected to lie in 68\% and 95\% of pseudo-experiments, respectively.  The thickness of the \zpssm\ theory curve represents the theoretical uncertainty from the PDF error set and $\alpha_S$, as well as the choice of PDF.
}
  \label{fig:combinedlimit_res}
\end{figure}

Figure~\ref{fig:combinedlimit_res} presents the expected and observed exclusion limits on \xbr\ at 95\%~CL for the combined 
dielectron and dimuon channels for the \zpssm\ search.  
The observed limit is within the $\pm2\sigma$ band of expected limits for all $M_{\zp}$. 
Figure~\ref{fig:combinedlimit_res} also contains the \zpssm\ theory band for \xbr. Its width represents the theoretical uncertainty, taking into account the following sources:
the PDF error set, the choice of PDF, and $\alpha_S$.
The value of $M_{\zp}$ at which the theory curve and the observed (expected) 95\%~CL limits on \xbr\
intersect is interpreted as the observed (expected) mass limit for the \zpssm boson, and corresponds
to \LimitCombined\ (\LimitCombinedExpected)~TeV.

\begin{figure}[t]
  \centering
\includegraphics[width=\columnwidth]{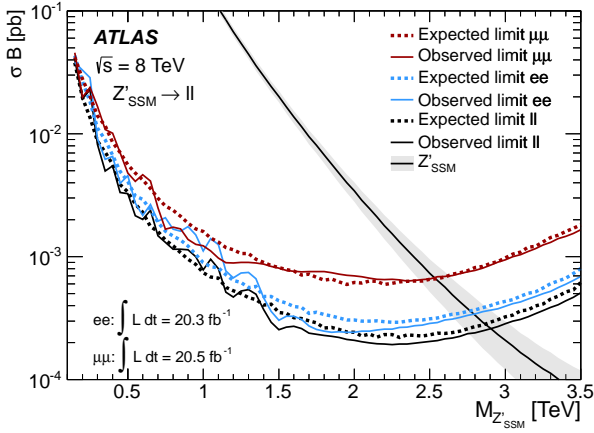}
  \caption{
Median expected (dashed line) and observed (solid line) 95\%~CL upper limits on cross-section times branching ratio (\xbr) for \zpssm\ production for the exclusive dimuon and dielectron channels, and for both channels combined.  The width of the \zpssm\ theory band represents the theoretical uncertainty from the PDF error set, the choice of PDF as well as $\alpha_S$.}
  \label{fig:Compilation_ZpSSM}
\end{figure}

A comparison of the combined limits on 
\xbr\ and those for the exclusive 
dielectron and dimuon channel is given in 
Figure~\ref{fig:Compilation_ZpSSM}. This demonstrates the contribution of 
each channel to the combined limit. As expected from Fig.~\ref{fig:SigEff_truem}, 
the larger values for $A \times \epsilon$ in addition to the better resolution in the 
dielectron channel results in a stronger limit
than in the dimuon channel. 
The observed (expected) \zpssm\ mass limit is \LimitElectron\ (\LimitElectronExpected)~TeV
in the dielectron channel, 
and \LimitMuon\ (\LimitMuonExpected)~TeV in the dimuon channel.

\begin{figure}[t]
  \centering
\includegraphics[width=\columnwidth]{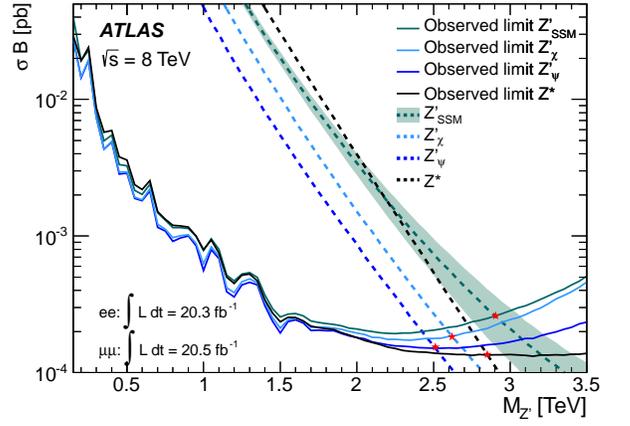}
  \caption{
Observed upper cross-section times branching ratio (\xbr) limits at 95\%~CL for \zpssm, \esix-motivated \zp\ and \zstar\ bosons using the combined dilepton channel.  In addition, theoretical cross-sections on \xbr\ are shown for the same models. The stars indicate the lower mass limits for each considered model.  The width of the \zpssm\ band represents the theoretical uncertainty from the PDF error set, the choice of PDF as well as $\alpha_S$.  The width of the \zpssm\ band applies to the \esix-motivated \zp\ curves as well.  
}
  \label{fig:Compilation_1Dlims}
\end{figure}

Figure~\ref{fig:Compilation_1Dlims} shows the observed \xbr\ exclusion limits at 95\%~CL
for the \zpssm, $\zp_{\chi}$, $\zp_{\psi}$ and 
\zstar\ signal searches.  
Here only observed limits are shown, as they are always very similar to the expected limits (see Fig.~\ref{fig:Compilation_ZpSSM}).
The theoretical \xbr\ 
of the boson for the \zpssm, two \esix-motivated Models and \zstar\ are also displayed.
The 95\%~CL limits on \xbr\ are used to set mass limits for each of the considered models.
Mass limits obtained for the \zpssm, \esix-motivated \zp\ and \zstar\ bosons are displayed in Table~\ref{tab:all_1D_limits_simple}.

As demonstrated in Fig.~\ref{fig:Compilation_1Dlims},
for lower values of $M_{\zp}$
the limit is driven primarily by the width of the signal
and gets stronger with decreasing width.  
At large $M_{\zp}$, 
the \xbr\ limit for a given \zp\ model worsens with increasing mass.
This weakening of the limit is due to the presence of the parton-luminosity tail in the \mll\ line shape.
The magnitude of this degradation is proportional to the size of the low-mass tail of the signal due to 
much higher background levels at low \mll\ compared to high \mll. 
All \zp\ models exhibit a parton-luminosity tail, the size of which increases with increasing natural width 
of the \zp\ resonance.
The tail is most pronounced for \zpssm, and least for $\zp_{\psi}$, in line with the different widths given in Table~\ref{tab:all_1D_limits_simple}.
Even though the width of the \zstar\ is similar to the width of the \zpssm, the 
tensor form of the coupling of the \zstar\ to fermions strongly 
suppresses parton luminosity effects. Limits on \xbr\ for the \zstar\ interpretation therefore do not worsen with increasing invariant mass.
Quantitatively, the observed \zpssm\ mass limit would increase from \LimitCombined~TeV to 2.95~TeV and 3.08~TeV, 
if the \zpchi\ and \zppsi\ boson signal templates, with smaller widths, were used.
If the \zstar\ boson template with negligible parton-luminosity tail but similar width were used instead of the \zpssm\ template, the 
observed limit would increase to 3.20~TeV.

\begin{table}[!hbt]
\caption{Observed and expected lower mass limits for \zp\ and \zstar\ bosons, using the corresponding signal template for a given model.}
\label{tab:all_1D_limits_simple}
\begin{center}
\begin{tabular}{c|c|c|c}
\hline
\hline
Model		&	Width     &       Observed  Limit &	Expected Limit \\
                        &             [\%]               &               [TeV]        &     [TeV]      \\
\hline
\zpssm\			&	3.0    &       2.90				&	2.87			 \\
$\zp_{\chi}$		&	1.2    &       2.62				&	2.60			 \\
$\zp_{\psi}$		&	0.5    &       2.51				&	2.46			 \\
\zstar\			&	3.4    &       \LimitCombinedZs 		&	\LimitCombinedExpectedZs \\
\hline
\hline
\end{tabular}
\end{center}
\end{table}


\subsection{\label{sec:mmzp}Limits on Minimal \zp\ bosons} 

Limits are also set in the \MM\ parameterization~\cite{Villadoro} of the \zp\ boson couplings introduced in Sec.~\ref{introMM}.  
Instead of using the predicted \xbr\ based on a fixed coupling to fermions as described in the previous section,
the new boson is characterized by two coupling parameters, \gbl\ and \gy.

\begin{figure}[t]
 \centering
  \includegraphics[width=\columnwidth]{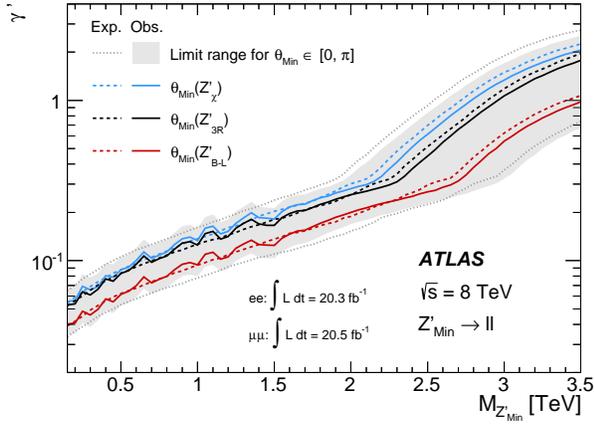}
 \caption{Expected (dotted and dashed lines) and observed (filled area and lines) limits at 95\%~CL on the strength of the \zp\ boson coupling relative to that of the SM $Z$ boson (\gammap) for the combined dielectron and dimuon channels as a function of the \zpMM\ mass in the \MM\ parameterization.  Limit curves are shown for three representative values of the mixing between the generators of the ($\mathrm{B-L}$) and the weak hypercharge Y gauge groups (\thetamin).  These are: ${\tan{\thetamin} = 0}$, ${\tan{\thetamin} = -2}$ and ${\tan{\thetamin} = -0.8}$, which correspond respectively to the \zpBL, \zpthreeR\ and \zpchi\ models at specific values of \gammap.  The region above each line is excluded. 
The gray band envelops all observed limit curves, which depend on the choice of ${\thetamin \in [0,\pi]}$. 
The corresponding expected limit curves 
are within the area delimited by the two dotted lines.
}
\label{fig:minimal_limits}
\end{figure}

\begin{figure}[t]
 \centering
  \includegraphics[width=\columnwidth]{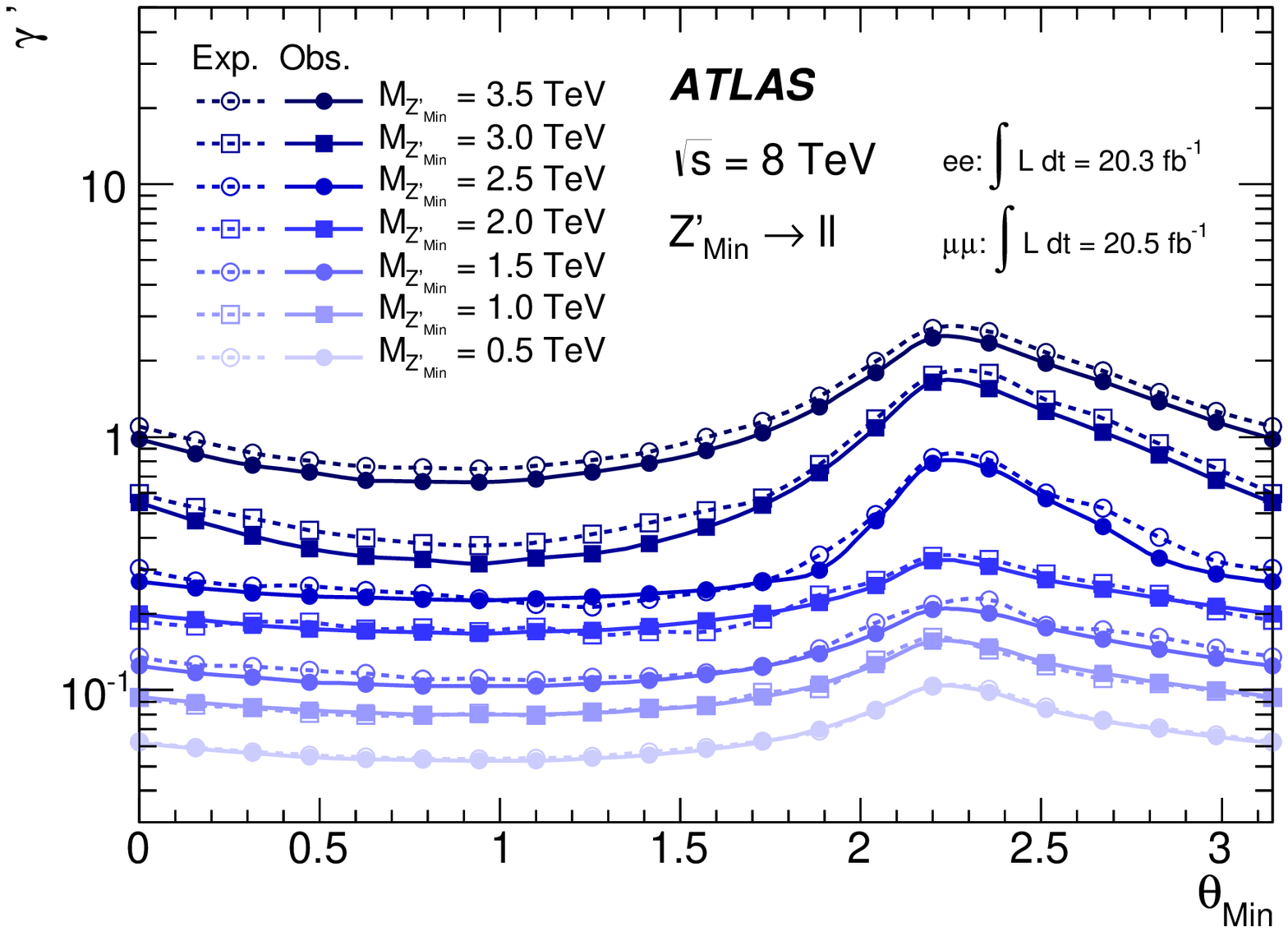}
 \caption{Expected (empty markers and dashed lines) and observed (filled markers and lines) limits at 95\%~CL on the strength of the \zp\ boson coupling relative to that of the SM $Z$ boson (\gammap) for the combined dielectron and dimuon channels as a function the mixing between the generators of the ($\mathrm{B-L}$) and the weak hypercharge Y gauge groups (\thetamin) in the \MM\ parameterization.  The limits are set for several representative values of the mass of the \zp\ boson, \Mmin.  The region above each line is excluded.
}
\label{fig:minimal_limits_vsTheta}
\end{figure}

For this analysis, the signal templates account for the dependence of the \zp\ boson width on 
\gammap\ and \thetamin, as well as the interference with SM \dy.
For a given value of \thetamin\ and for each tested \zp\ mass, 
dilepton invariant mass templates are created with various \gammap\ values between 0.005 and 4.
The templates at these chosen values of \gammap\ are interpolated to other values of \gammap\  by using a smooth interpolating function in each dilepton
mass bin. The parameter of interest in the likelihood analysis is \gammap\  for specific values of \thetamin\ and the \zp\ boson mass, \Mmin. 
Systematic uncertainties are included in the analysis analogously to the computation of \xbr\ limits described above. 
Limits at 95\%~CL are set on the relative coupling strength \gammap\ as a function of the \zpMM\ boson mass, as shown in Fig.~\ref{fig:minimal_limits}. 
Figure~\ref{fig:minimal_limits_vsTheta} contains limits at 95\%~CL on \gammap\ versus \thetamin\ for several representative values of \Mmin.
The strongest and weakest limits are found for $\thetamin =  0.96$ and $\thetamin = 2.27$, respectively. 
The limits depend heavily on the \zp\ branching ratio to dileptons, which in turn depends on \thetamin\ as the choice of this parameter influences the \zp\ couplings.
For \Mmin\ significantly above the \ttbar\ production threshold, the sum of \zp\ branching ratios to electron and muon pairs ranges from 4.6\% to 32\%.

\subsection{\label{sec:gslimits}Limits on spin-2 graviton excitations in Randall--Sundrum models}

\begin{figure}
  \centering
  \includegraphics[width=\columnwidth]{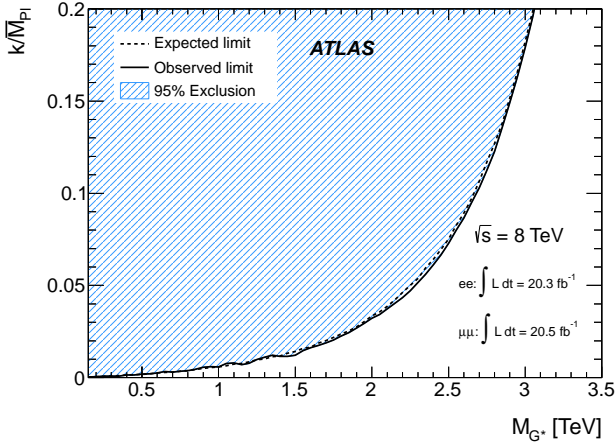}
  \caption{Expected and observed 95\%~CL limits in the plane of the coupling strength of the Randall--Sundrum \gstar\ to SM particles (\kovermb) versus \gstar\ mass for the combination of the dielectron and dimuon channels. The region above the curve is excluded at 95\%~CL.
}
  \label{fig:2Dlimit_RS}
\end{figure}

The phenomenology of RS models is characterized by the \gstar\ mass and \kovermb.
Limits at 95\%~CL on $\xbr (\gstar \to \ell^+\ell^-)$ are obtained and compared to the theoretical \xbr\
assuming values of \kovermb\ less than 0.2.
These results are used to set limits in the plane of \kovermb\ versus \gstar\ mass, as illustrated in Fig.~\ref{fig:2Dlimit_RS} for the combined dilepton channel. 
Mass limits for five of the \kovermb\ values used are given in Table~\ref{tab:combinedLimitsG_coupl}.

\begin{table}[!h]
\caption{
Observed and expected 95\% CL lower limits on the mass of the \gstar\ with varying coupling \kovermb. 
The two lepton channels are combined.
}
\label{tab:combinedLimitsG_coupl}
\begin{center}
\begin{tabular}{l|ccccc}
\hline
\hline
\rule{0pt}{1.05em}
\kovermb            &  0.01 & 0.03  & 0.05 & 0.1 & 0.2\\
\hline			 	 	 	 	  	 
Observed limit on $M_{\gstar}$ [TeV] & \LimitCombinedGOne & \LimitCombinedGThree  & \LimitCombinedGFive  & \LimitCombinedG & \LimitCombinedGTwo\\
Expected limit on $M_{\gstar}$ [TeV] & \LimitCombinedGExpOne & \LimitCombinedGExpThree  & \LimitCombinedGExpFive  & \LimitCombinedExpectedG  & \LimitCombinedExpectedGTwo\\
\hline
\hline
\end{tabular}
\end{center}
\end{table}


\subsection{\label{sec:qbh}Limits on quantum black hole models }

\begin{figure}[b]
  \centering
  \includegraphics[width=\columnwidth]{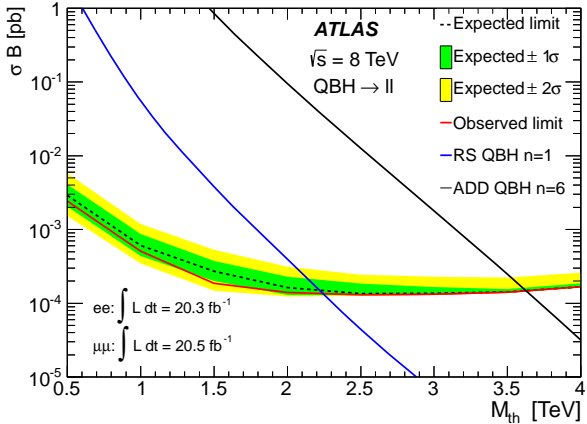}
  \caption{Expected and observed 95\%~CL upper limits on cross-section times branching ratio (\xbr) for quantum black hole production in the extra-dimensional model proposed by Arkani-Hamed, Dimopoulos and Dvali (ADD)
and Randall--Sundrum (RS) for the combined dilepton channel.  
}
  \label{fig:qbh_both_limit_comb}
\end{figure}

Upper limits at 95\%~CL on \xbr\ are set as a function of $M_{\mathrm{th}}$, assuming 
a signal according to both the RS and ADD models. While the two models predict different mass distributions, 
using the same \xbr\ limit curve for each (as in Fig.~\ref{fig:qbh_both_limit_comb}) affects the mass limits obtained by only 1\%.  
The observed lower limits on $M_{\mathrm{th}}$ for the combination of the two dilepton channels are 3.65~TeV for the ADD model and 2.24~TeV for the RS model.

\subsection{\label{sec:MWTlimits}Limits on Minimal Walking Technicolor}

\begin{figure}[t]
 \centering
  \includegraphics[width=\columnwidth]{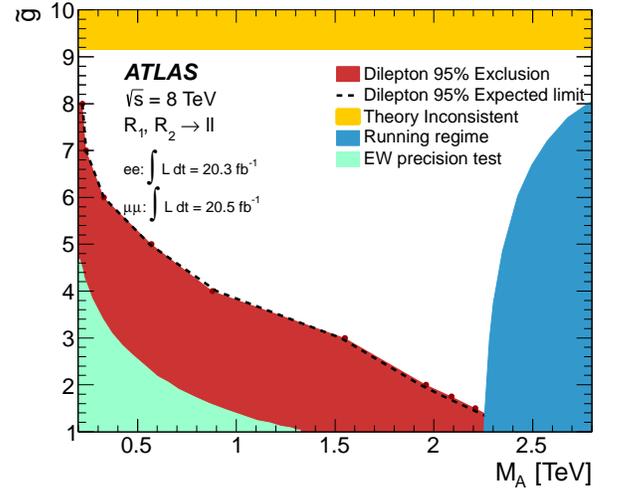}
 \caption{
Exclusion contours at 95\%~CL in the plane of the Minimal Walking Technicolor parameter space defined by the bare axial-vector mass versus the strength of the spin-1 resonance interaction ($M_A$,
$\tilde{g}$). Electroweak precision measurements exclude the green area in the bottom left corner. 
The requirement to stay in the walking regime excludes the blue area in the right corner. 
The red area (black dashed line) shows the observed (expected) exclusion for both channels combined.
The upper region is excluded due to non-real axial and axial-vector decay constants.
}
\label{fig:MWT_2D}
\end{figure}

The MWT model, introduced in Sec.~\ref{introMWT}, is tested by searching for technimeson resonances.
Limits on \xbr\ are set at 95\%~CL as a function of $M_{\Rone}$ for $\gtilde=2$.  
Electroweak precision data, a requirement to stay in the walking technicolor regime and constraints 
from requiring real-valued physical decay constants exclude a portion of the \gtilde\ versus $M_A$ plane, as shown in Fig.~\ref{fig:MWT_2D}.  
By combining these factors and the 95\%~CL limits that are set, all possible $M_A$ masses are excluded for $\gtilde$ less than $\approx 1.4$. 
Limits on $M_{\Rone}$ for various values of $\gtilde$ are given in Table~\ref{tab:limits_MWTC}.

\begingroup
\squeezetable
\begin{table}[h]
\caption{
Combined 95\%~CL observed and expected lower mass limits on $M_{\Rone}$ and $M_A$ (Minimal Walking Technicolor model) for various values of $\gtilde$.
}
\label{tab:limits_MWTC}
\begin{center}
\begin{tabular}{l|cccccccc}
\hline
\hline
~~~~~~~~~~~~~~\gtilde            &  1.5	 & 2	  & 3	 & 4	 & 5	& 6	& 7	& 8	\\
\hline			 	 	 	 	  	 
Observed limit $M_{\Rone}$ [TeV] & \LimCombMWTCOneFive 		& \LimCombMWTCTwo  		& \LimCombMWTCThree  	& \LimCombMWTCFour 		& \LimCombMWTCFive		& \LimCombMWTCSix		& \LimCombMWTCSeven		& \LimCombMWTCEight		\\
Expected limit $M_{\Rone}$ [TeV] & \LimCombExpMWTCOneFive 	& \LimCombExpMWTCTwo  	& \LimCombExpMWTCThree  	& \LimCombExpMWTCFour  	& \LimCombExpMWTCFive	& \LimCombExpMWTCSix	& \LimCombExpMWTCSeven	& \LimCombExpMWTCEight	\\
\hline
Observed limit $M_A$ [TeV] & 2.21 	& 1.96  	& 1.55  	& 0.88 	& 0.57	& 0.33	& 0.24	& 0.22		\\
Expected limit $M_A$ [TeV] & 2.18 	& 1.93  	& 1.53  	& 0.90  & 0.56	& 0.33	& 0.24	& 0.22		\\
\hline
\hline
\end{tabular}
\end{center}
\end{table}
\endgroup



\section{Conclusions}

In conclusion, the \mbox{ATLAS} detector at the Large Hadron Collider was used to search for 
resonances decaying to dielectron or dimuon final states at masses above the pole mass of the $Z$ boson,
using \LumiElectronsInFb\ of proton--proton 
collision data collected in 2012 at $\rts = 8$~TeV in the dielectron channel, and \LumiMuonsInFb\ in the dimuon channel. 
The observed invariant mass spectrum is consistent with the Standard Model expectation. 
Limits are set on signal cross-section times branching fraction for a variety of physics scenarios beyond the Standard Model.
For the benchmark \zpssm\ boson with a mass of 2.5~TeV, the expected cross-section limit improved approximately fivefold in comparison to 
the previous \mbox{ATLAS} publication, which used $\rts = 7$~TeV data collected in 2011.
The limit on the mass of the benchmark \zpssm\ signal improved from 2.22~TeV to \LimitCombined~TeV, and mass limits of \LimitCombinedPsi$-$\LimitCombinedChi~TeV are set for various
\esix-motivated \zp~bosons.  For \zstar~bosons, the mass limit is \LimitCombinedZs~TeV, and the limit on the mass of the \gstar\ in the Randall--Sundrum model with coupling parameter \kovermb\ equal to 0.1 is \LimitCombinedG~TeV.  Experimental limits are also set on \MM\ and on a Minimal Walking Technicolor 
model with a composite Higgs boson.  The limits set on the production threshold of quantum black holes are 3.65~TeV for the extra-dimensional model proposed by Arkani-Hamed, Dimopoulos and Dvali 
and 2.24~TeV for the Randall--Sundrum model.  For all but those on quantum black hole production, the limits presented are the most stringent to date.



\section{Acknowledgements}

We thank T. Hapola for implementing the Minimal Walking Technicolor model using MadGraph 
to generate the signal and for his help with acceptance studies. 
The limits shown in Section XII were calculated using computing resources provided by the 
Argonne Leadership Computing Facility and the National Energy Research Scientific Computing Center.

We thank CERN for the very successful operation of the LHC, as well as the
support staff from our institutions without whom ATLAS could not be
operated efficiently.

We acknowledge the support of ANPCyT, Argentina; YerPhI, Armenia; ARC,
Australia; BMWF and FWF, Austria; ANAS, Azerbaijan; SSTC, Belarus; CNPq and FAPESP,
Brazil; NSERC, NRC and CFI, Canada; CERN; CONICYT, Chile; CAS, MOST and NSFC,
China; COLCIENCIAS, Colombia; MSMT CR, MPO CR and VSC CR, Czech Republic;
DNRF, DNSRC and Lundbeck Foundation, Denmark; EPLANET, ERC and NSRF, European Union;
IN2P3-CNRS, CEA-DSM/IRFU, France; GNSF, Georgia; BMBF, DFG, HGF, MPG and AvH
Foundation, Germany; GSRT and NSRF, Greece; ISF, MINERVA, GIF, I-CORE and Benoziyo Center,
Israel; INFN, Italy; MEXT and JSPS, Japan; CNRST, Morocco; FOM and NWO,
Netherlands; BRF and RCN, Norway; MNiSW and NCN, Poland; GRICES and FCT, Portugal; MNE/IFA, Romania; MES of Russia and ROSATOM, Russian Federation; JINR; MSTD,
Serbia; MSSR, Slovakia; ARRS and MIZ\v{S}, Slovenia; DST/NRF, South Africa;
MINECO, Spain; SRC and Wallenberg Foundation, Sweden; SER, SNSF and Cantons of
Bern and Geneva, Switzerland; NSC, Taiwan; TAEK, Turkey; STFC, the Royal
Society and Leverhulme Trust, United Kingdom; DOE and NSF, United States of
America.

The crucial computing support from all WLCG partners is acknowledged
gratefully, in particular from CERN and the ATLAS Tier-1 facilities at
TRIUMF (Canada), NDGF (Denmark, Norway, Sweden), CC-IN2P3 (France),
KIT/GridKA (Germany), INFN-CNAF (Italy), NL-T1 (Netherlands), PIC (Spain),
ASGC (Taiwan), RAL (UK) and BNL (USA) and in the Tier-2 facilities
worldwide.

\bibliographystyle{atlasBibStyleWoTitle}
\bibliography{Zp}{}

\providecommand{\href}[2]{#2}\begingroup\raggedright\begin{thebibliography}{10}

\bibitem{atlas:detector}
{ATLAS} Collaboration,
  \href{http://dx.doi.org/10.1088/1748-0221/3/08/S08003}{JINST {\bfseries 3}
  (2008) S08003}.

\bibitem{London:1986dk}
D.~London and J.~L. Rosner,
\href{http://dx.doi.org/10.1103/PhysRevD.34.1530}{Phys.~Rev. {\bfseries D34}
  (1986) 1530}.

\bibitem{Langacker:2008yv}
P.~Langacker,
  \href{http://dx.doi.org/10.1103/RevModPhys.81.1199}{Rev.~Mod.~Phys.
  {\bfseries 81} (2009) 1199},
\href{http://arxiv.org/abs/0801.1345}{{\ttfamily arXiv:0801.1345 [hep-ph]}}.

\bibitem{Villadoro}
E.~Salvioni, G.~Villadoro, and F.~Zwirner,
  \href{http://dx.doi.org/10.1088/1126-6708/2009/11/068}{JHEP {\bfseries 0911}
  (2009) 068},
\href{http://arxiv.org/abs/0909.1320}{{\ttfamily arXiv:0909.1320 [hep-ph]}}.

\bibitem{atlasHiggs}
{ATLAS} Collaboration,
  \href{http://dx.doi.org/10.1016/j.physletb.2012.08.020}{Phys. Lett.
  {\bfseries B716} (2012) 1},
\href{http://arxiv.org/abs/1207.7214}{{\ttfamily arXiv:1207.7214 [hep-ex]}}.

\bibitem{cmsHiggs}
{CMS} Collaboration,
  \href{http://dx.doi.org/10.1016/j.physletb.2012.08.021}{Phys. Lett.
  {\bfseries B716} (2012) 30},
\href{http://arxiv.org/abs/1207.7235}{{\ttfamily arXiv:1207.7235 [hep-ex]}}.

\bibitem{wzstar}
M.~V. Chizhov, V.~A. Bednyakov, and J.~A. Budagov,
  \href{http://dx.doi.org/10.1134/S1063778808120107}{Physics of Atomic Nuclei
  {\bfseries 71} (2008) 2096}, \href{http://arxiv.org/abs/0801.4235}{{\ttfamily
  arXiv:0801.4235 [hep-ph]}}.

\bibitem{wzstar_motivate}
M.~V. Chizhov and G.~Dvali,
  \href{http://dx.doi.org/10.1016/j.physletb.2011.08.056}{Phys.~Lett.
  {\bfseries B703} (2011) 593},
\href{http://arxiv.org/abs/0908.0924}{{\ttfamily arXiv:0908.0924 [hep-ph]}}.

\bibitem{wzstar_refmod}
M.~V. Chizhov, \href{http://dx.doi.org/10.1134/S1547477111060045}{Phys.~Part.
  Nucl. Lett. {\bfseries 8} (2011) 512},
\href{http://arxiv.org/abs/1005.4287}{{\ttfamily arXiv:1005.4287 [hep-ph]}}.

\bibitem{wzstar_refmod2}
M.~V. Chizhov, V.~A. Bednyakov, and J.~A. Budagov, Nuovo Cimento {\bfseries
  C33} (2010) 343, \href{http://arxiv.org/abs/1005.2728}{{\ttfamily
  arXiv:1005.2728 [hep-ph]}}.

\bibitem{RS}
L.~Randall and R.~Sundrum, Phys.~Rev.~Lett. {\bfseries 83} (1999) 3370,
  \href{http://arxiv.org/abs/hep-ph/9905221}{{\ttfamily arXiv:hep-ph/9905221}}.

\bibitem{ref:Meade_QBH}
P.~Meade and L.~Randall,
  \href{http://dx.doi.org/10.1088/1126-6708/2008/05/003}{JHEP {\bfseries 0805}
  (2008) 003}, \href{http://arxiv.org/abs/0708.3017}{{\ttfamily arXiv:0708.3017
  [hep-ph]}}.

\bibitem{TC3}
F.~Sannino and K.~Tuominen, Phys.~Rev. {\bfseries D71} (2005) 051901,
  \href{http://arxiv.org/abs/hep-ph/0405209}{{\ttfamily arXiv:hep-ph/0405209}}.

\bibitem{TC4}
D.~D. Dietrich, F.~Sannino, and K.~Tuominen, Phys.~Rev. {\bfseries D72} (2005)
  055001, \href{http://arxiv.org/abs/hep-ph/0505059}{{\ttfamily
  arXiv:hep-ph/0505059}}.

\bibitem{TC5}
R.~Foadi, M.~T. Frandsen, T.~A. Ryttov, and F.~Sannino,
  \href{http://dx.doi.org/10.1103/PhysRevD.76.055005}{Phys.~Rev. {\bfseries
  D76} (2007) 055005}, \href{http://arxiv.org/abs/0706.1696}{{\ttfamily
  arXiv:0706.1696 [hep-ph]}}.

\bibitem{Foadi:2012bb}
R.~Foadi, M.~T. Frandsen, and F.~Sannino,
  \href{http://dx.doi.org/10.1103/PhysRevD.87.095001}{Phys. Rev. {\bfseries
  D87} (2013) 095001},
\href{http://arxiv.org/abs/1211.1083}{{\ttfamily arXiv:1211.1083 [hep-ph]}}.

\bibitem{zprime_2011_paper}
{ATLAS Collaboration}, \href{http://dx.doi.org/10.1007/JHEP11(2012)138}{JHEP
  {\bfseries 1211} (2012) 138},
  \href{http://arxiv.org/abs/1209.2535}{{\ttfamily arXiv:1209.2535 [hep-ex]}}.

\bibitem{Chatrchyan201363}
{CMS} Collaboration,
  \href{http://dx.doi.org/10.1016/j.physletb.2013.02.003}{Phys. Lett.
  {\bfseries B720} (2013) 63},
\href{http://arxiv.org/abs/1212.6175}{{\ttfamily arXiv:1212.6175 [hep-ex]}}.

\bibitem{Abazov:2010ti}
{D0 Collaboration, V. M. Abazov et al.},
  \href{http://dx.doi.org/10.1016/j.physletb.2010.10.059}{Phys.~Lett.
  {\bfseries B695} (2011) 88},
\href{http://arxiv.org/abs/1008.2023}{{\ttfamily arXiv:1008.2023 [hep-ex]}}.

\bibitem{CDF:Zpmumu}
{CDF Collaboration, T. Aaltonen et al.},
  \href{http://dx.doi.org/10.1103/PhysRevLett.106.121801}{Phys.~Rev.~Lett.
  {\bfseries 106} (2011) 121801},
\href{http://arxiv.org/abs/1101.4578}{{\ttfamily arXiv:1101.4578 [hep-ex]}}.

\bibitem{Abbiendi:2003dh}
{OPAL Collaboration, G. Abbiendi et al.},
  \href{http://dx.doi.org/10.1140/epjc/s2004-01595-9}{Eur.~Phys.~J. {\bfseries
  C33} (2004) 173}, \href{http://arxiv.org/abs/hep-ex/0309053}{{\ttfamily
  arXiv:hep-ex/0309053}}.

\bibitem{Abdallah:2005ph}
{DELPHI Collaboration, J. Abdallah et al.}, Eur.~Phys.~J. {\bfseries C45}
  (2006) 589, \href{http://arxiv.org/abs/hep-ex/0512012}{{\ttfamily
  arXiv:hep-ex/0512012}}.

\bibitem{Achard:2005nb}
{L3 Collaboration, P. Achard et al.}, Eur.~Phys.~J. {\bfseries C47} (2006) 1,
  \href{http://arxiv.org/abs/hep-ex/0603022}{{\ttfamily arXiv:hep-ex/0603022}}.

\bibitem{Schael:2006wu}
{ALEPH Collaboration, S. Schael et al.}, Eur.~Phys.~J. {\bfseries C49} (2007)
  411, \href{http://arxiv.org/abs/hep-ex/0609051}{{\ttfamily
  arXiv:hep-ex/0609051}}.

\bibitem{Langacker:2009su}
P.~Langacker,
\href{http://arxiv.org/abs/0911.4294}{{\ttfamily arXiv:0911.4294 [hep-ph]}}.

\bibitem{Dittmar:2003ir}
M.~Dittmar, A.-S. Nicollerat, and A.~Djouadi, Phys.~Lett. {\bfseries B583}
  (2004) 111, \href{http://arxiv.org/abs/hep-ph/0307020}{{\ttfamily
  arXiv:hep-ph/0307020}}.

\bibitem{Accomando:2010fz}
E.~Accomando {et~al.},
  \href{http://dx.doi.org/10.1103/PhysRevD.83.075012}{Phys.~Rev. {\bfseries
  D83} (2011) 075012}, \href{http://arxiv.org/abs/1010.6058}{{\ttfamily
  arXiv:1010.6058 [hep-ph]}}.

\bibitem{Senjanovic:1975rk}
G.~Senjanovic and R.~N. Mohapatra,
\href{http://dx.doi.org/10.1103/PhysRevD.12.1502}{Phys.~Rev. {\bfseries D12}
  (1975) 1502}.

\bibitem{Mohapatra:1974hk}
R.~N. Mohapatra and J.~C. Pati,
\href{http://dx.doi.org/10.1103/PhysRevD.11.566}{Phys. Rev. {\bfseries D11}
  (1975) 566}.

\bibitem{Basso:2008iv}
L.~Basso, A.~Belyaev, S.~Moretti, and C.~H. Shepherd-Themistocleous,
  \href{http://dx.doi.org/10.1103/PhysRevD.80.055030}{Phys. Rev. {\bfseries
  D80} (2009) 055030},
\href{http://arxiv.org/abs/0812.4313}{{\ttfamily arXiv:0812.4313 [hep-ph]}}.

\bibitem{Davoudiasl2000}
H.~Davoudiasl, J.~L. Hewett, and T.~G. Rizzo,
  \href{http://dx.doi.org/10.1103/PhysRevLett.84.2080}{Phys.~Rev.~Lett.
  {\bfseries 84} (2000) 2080},
  \href{http://arxiv.org/abs/hep-ph/9909255}{{\ttfamily arXiv:hep-ph/9909255}}.

\bibitem{ref:Dimopoulos_QBH}
D.~Dimopoulos and G.~Landsberg,
  \href{http://dx.doi.org/10.1103/PhysRevLett.87.161602}{Phys. Rev. Lett.
  {\bfseries 87} (2001) 161602},
  \href{http://arxiv.org/abs/hep-ph/0106295}{{\ttfamily arXiv:hep-ph/0106295}}.

\bibitem{ref:Giddings_QBH}
S.~B. Giddings and S.~Thomas,
  \href{http://dx.doi.org/10.1103/PhysRevD.65.056010}{Phys. Rev. {\bfseries
  D65} (2002) 056010}, \href{http://arxiv.org/abs/hep-ph/0106219}{{\ttfamily
  arXiv:hep-ph/0106219}}.

\bibitem{ref:GingrichMartell_QBH}
D.~M. Gingrich and K.~Martell,
  \href{http://dx.doi.org/10.1103/PhysRevD.78.115009}{Phys. Rev. {\bfseries
  D78} (2008) 115009}, \href{http://arxiv.org/abs/0808.2512}{{\ttfamily
  arXiv:0808.2512 [hep-ph]}}.

\bibitem{ref:Calmet_QBH}
X.~Calmet, W.~Gong, and S.~D.~H. Hsu,
  \href{http://dx.doi.org/10.1016/j.physletb.2008.08.011}{Phys. Lett.
  {\bfseries B668} (2008) 20}, \href{http://arxiv.org/abs/0806.4605}{{\ttfamily
  arXiv:0806.4605 [hep-ph]}}.

\bibitem{ArkaniHamed:1998rs}
N.~Arkani-Hamed, S.~Dimopoulos, and G.~Dvali,
  \href{http://dx.doi.org/10.1016/S0370-2693(98)00466-3}{Phys. Lett. {\bfseries
  B429} (1998) 263}, \href{http://arxiv.org/abs/hep-ph/9803315}{{\ttfamily
  arXiv:hep-ph/9803315}}.

\bibitem{ref:Gingrich_QBH}
D.~M. Gingrich, \href{http://dx.doi.org/10.1088/0954-3899/37/10/105008}{J.
  Phys. G: Nucl. Part. Phys. {\bfseries 37} (2010) 105008},
  \href{http://arxiv.org/abs/0912.0826}{{\ttfamily arXiv:0912.0826 [hep-ph]}}.

\bibitem{ATLAS_qbh_gj}
{ATLAS Collaboration},
  \href{http://dx.doi.org/10.1016/j.physletb.2013.12.029}{Phys. Lett.
  {\bfseries B728} (2014) 562},
\href{http://arxiv.org/abs/1309.3230}{{\ttfamily arXiv:1309.3230 [hep-ex]}}.

\bibitem{Aad:2013gma}
{ATLAS} Collaboration,
\href{http://arxiv.org/abs/1311.2006}{{\ttfamily arXiv:1311.2006 [hep-ex]}}.

\bibitem{cmsDijet_qbh}
{CMS} Collaboration, \href{http://dx.doi.org/10.1007/JHEP07(2013)178}{JHEP
  {\bfseries 07} (2013) 178},
\href{http://arxiv.org/abs/1303.5338}{{\ttfamily arXiv:1303.5338 [hep-ex]}}.

\bibitem{blackmax}
D.-C. Dai {et~al.}, \href{http://dx.doi.org/10.1103/PhysRevD.77.076007}{Phys.
  Rev. {\bfseries D77} (2008) 076007},
\href{http://arxiv.org/abs/0711.3012}{{\ttfamily arXiv:0711.3012 [hep-ph]}}.

\bibitem{ATLAS_qbh_dijet}
{ATLAS Collaboration}, \href{http://dx.doi.org/10.1007/JHEP01(2013)029}{JHEP
  {\bfseries 1301} (2013) 029},
  \href{http://arxiv.org/abs/1210.1718}{{\ttfamily arXiv:1210.1718 [hep-ex]}}.

\bibitem{Andersen:2011nk}
J.~R. Andersen, T.~Hapola, and F.~Sannino,
  \href{http://dx.doi.org/10.1103/PhysRevD.85.055017}{Phys. Rev. {\bfseries
  D85} (2012) 055017},
\href{http://arxiv.org/abs/1105.1433}{{\ttfamily arXiv:1105.1433 [hep-ph]}}.

\bibitem{Belyaev:2008yj}
A.~Belyaev {et~al.}, \href{http://dx.doi.org/10.1103/PhysRevD.79.035006}{Phys.
  Rev. {\bfseries D79} (2009) 035006},
\href{http://arxiv.org/abs/0809.0793}{{\ttfamily arXiv:0809.0793 [hep-ph]}}.

\bibitem{atlas_trigger}
{ATLAS} Collaboration,
  \href{http://dx.doi.org/10.1140/epjc/s10052-011-1849-1}{Eur.~Phys.~J.
  {\bfseries C72} (2012) 1849},
  \href{http://arxiv.org/abs/1110.1530}{{\ttfamily arXiv:1110.1530 [hep-ex]}}.

\bibitem{atlas:egamma_perf2011}
{{ATLAS}} Collaboration,
\href{http://arxiv.org/abs/1404.2240}{{\ttfamily arXiv:1404.2240 [hep-ex]}}.

\bibitem{Alioli:2010xd}
S.~Alioli, P.~Nason, C.~Oleari, and E.~Re,
  \href{http://dx.doi.org/10.1007/JHEP06(2010)043}{JHEP {\bfseries 1006} (2010)
  043},
\href{http://arxiv.org/abs/1002.2581}{{\ttfamily arXiv:1002.2581 [hep-ph]}}.

\bibitem{CT10}
H.-L. Lai {et~al.},
  \href{http://dx.doi.org/10.1103/PhysRevD.82.074024}{Phys.~Rev. {\bfseries
  D82} (2010) 074024}, \href{http://arxiv.org/abs/1007.2241}{{\ttfamily
  arXiv:1007.2241 [hep-ph]}}.

\bibitem{pythia8}
T.~Sj{\"o}strand, S.~Mrenna, and P.~Z. Skands,
  \href{http://dx.doi.org/10.1016/j.cpc.2008.01.036}{Comput.~Phys.~Commun.
  {\bfseries 178} (2008) 852}, \href{http://arxiv.org/abs/0710.3820}{{\ttfamily
  arXiv:0710.3820 [hep-ph]}}.

\bibitem{fsr_ref}
P.~Golonka and Z.~W\c{a}s,
  \href{http://dx.doi.org/10.1140/epjc/s2005-02396-4}{Eur.~Phys.~J. {\bfseries
  C45} (2006) 97}, \href{http://arxiv.org/abs/hep-ph/0506026}{{\ttfamily
  arXiv:hep-ph/0506026}}.

\bibitem{atlas:sim}
{ATLAS Collaboration},
  \href{http://dx.doi.org/10.1140/epjc/s10052-010-1429-9}{Eur.~Phys.~J.
  {\bfseries C70} (2010) 823},
\href{http://arxiv.org/abs/1005.4568}{{\ttfamily arXiv:1005.4568
  [physics.ins-det]}}.

\bibitem{geant}
{GEANT4 Collaboration, S. Agostinelli et al.},
  \href{http://dx.doi.org/10.1016/S0168-9002(03)01368-8}{Nucl. Instrum. Methods
  Phys. Res., Sect. A {\bfseries 506} (2003) 250}.

\bibitem{fewz}
K.~Melnikov and F.~Petriello,
  \href{http://dx.doi.org/10.1103/PhysRevD.74.114017}{Phys.~Rev. {\bfseries
  D74} (2006) 114017}, \href{http://arxiv.org/abs/hep-ph/0609070}{{\ttfamily
  arXiv:hep-ph/0609070}}.

\bibitem{fewz1}
{Y. Li and F. Petriello},
  \href{http://dx.doi.org/10.1103/PhysRevD.86.094034}{Phys.~Rev. {\bfseries
  D86} (2012) 094034}, \href{http://arxiv.org/abs/1208.5967}{{\ttfamily
  arXiv:1208.5967 [hep-ph]}}.

\bibitem{mstw}
A.~D. Martin, W.~J. Stirling, R.~S. Thorne, and G.~Watt,
  \href{http://dx.doi.org/10.1140/epjc/s10052-009-1072-5}{Eur. Phys. J.
  {\bfseries C63} (2009) 189},
\href{http://arxiv.org/abs/0901.0002}{{\ttfamily arXiv:0901.0002 [hep-ph]}}.

\bibitem{MRST2004QED}
A.~D. Martin, R.~G. Roberts, W.~J. Stirling, and R.~S. Thorne, Eur.~Phys.~J.
  {\bfseries C39} (2005) 155,
  \href{http://arxiv.org/abs/hep-ph/0411040}{{\ttfamily arXiv:hep-ph/0411040}}.

\bibitem{Bardin:2012jk}
D.~Bardin {et~al.}, \href{http://dx.doi.org/10.1134/S002136401217002X}{JETP
  Lett. {\bfseries 96} (2012) 285--289},
\href{http://arxiv.org/abs/1207.4400}{{\ttfamily arXiv:1207.4400 [hep-ph]}}.

\bibitem{Bondarenko:2013nu}
S.~G. Bondarenko and A.~A. Sapronov,
  \href{http://dx.doi.org/10.1016/j.cpc.2013.05.010}{Comput. Phys. Commun.
  {\bfseries 184} (2013) 2343},
\href{http://arxiv.org/abs/1301.3687}{{\ttfamily arXiv:1301.3687 [hep-ph]}}.

\bibitem{madgraph5}
J.~Alwall {et~al.}, \href{http://dx.doi.org/10.1007/JHEP06(2011)128}{JHEP
  {\bfseries 1106} (2011) 128},
  \href{http://arxiv.org/abs/1106.0522}{{\ttfamily arXiv:1106.0522 [hep-ph]}}.

\bibitem{ewrad}
U.~Baur, Phys.~Rev. {\bfseries D75} (2007) 013005,
  \href{http://arxiv.org/abs/hep-ph/0611241}{{\ttfamily arXiv:hep-ph/0611241}}.

\bibitem{herwig}
G.~Corcella {et~al.},
  \href{http://dx.doi.org/10.1088/1126-6708/2001/01/010}{JHEP {\bfseries 0101}
  (2001) 010}, \href{http://arxiv.org/abs/hep-ph/0011363}{{\ttfamily
  arXiv:hep-ph/0011363}}.

\bibitem{Corcella:2002jc}
G.~Corcella {et~al.}, \href{http://arxiv.org/abs/hep-ph/0210213}{{\ttfamily
  arXiv:hep-ph/0210213}}.

\bibitem{Pumplin:2002vw}
J.~Pumplin {et~al.},
  \href{http://dx.doi.org/10.1088/1126-6708/2002/07/012}{JHEP {\bfseries 0207}
  (2002) 012}, \href{http://arxiv.org/abs/hep-ph/0201195}{{\ttfamily
  arXiv:hep-ph/0201195}}.

\bibitem{Campbell:1999mcfm}
J.~M. Campbell and R.~K. Ellis,
  \href{http://dx.doi.org/10.1103/PhysRevD.60.113006}{Phys.~Rev. {\bfseries
  D60} (1999) 113006}, \href{http://arxiv.org/abs/hep-ph/9905386}{{\ttfamily
  arXiv:hep-ph/9905386}}.

\bibitem{mcatnlo}
S.~Frixione and B.~R. Webber,
  \href{http://dx.doi.org/10.1088/1126-6708/2002/06/029}{JHEP {\bfseries 0206}
  (2002) 029}, \href{http://arxiv.org/abs/hep-ph/0204244}{{\ttfamily
  arXiv:hep-ph/0204244}}.

\bibitem{Frixione:2003ei}
S.~Frixione, P.~Nason, and B.~R. Webber,
  \href{http://dx.doi.org/10.1088/1126-6708/2003/08/007}{JHEP {\bfseries 0308}
  (2003) 007},
\href{http://arxiv.org/abs/hep-ph/0305252}{{\ttfamily arXiv:hep-ph/0305252
  [hep-ph]}}.

\bibitem{Frixione:2008yi}
S.~Frixione {et~al.},
  \href{http://dx.doi.org/10.1088/1126-6708/2008/07/029}{JHEP {\bfseries 0807}
  (2008) 029},
\href{http://arxiv.org/abs/0805.3067}{{\ttfamily arXiv:0805.3067 [hep-ph]}}.

\bibitem{Cacciari:2011}
M.~Cacciari {et~al.},
  \href{http://dx.doi.org/10.1016/j.physletb.2012.03.013}{Phys. Lett.
  {\bfseries B710} (2012) 612},
\href{http://arxiv.org/abs/1111.5869}{{\ttfamily arXiv:1111.5869 [hep-ph]}}.

\bibitem{Baernreuther:2012ws}
P.~B{\"a}rnreuther, M.~Czakon, and A.~Mitov,
  \href{http://dx.doi.org/10.1103/PhysRevLett.109.132001}{Phys. Rev. Lett.
  {\bfseries 109} (2012) 132001},
\href{http://arxiv.org/abs/1204.5201}{{\ttfamily arXiv:1204.5201 [hep-ph]}}.

\bibitem{Czakon:2012zr}
M.~Czakon and A.~Mitov, \href{http://dx.doi.org/10.1007/JHEP12(2012)054}{JHEP
  {\bfseries 1212} (2012) 054},
\href{http://arxiv.org/abs/1207.0236}{{\ttfamily arXiv:1207.0236 [hep-ph]}}.

\bibitem{Czakon:2012pz}
M.~Czakon and A.~Mitov, \href{http://dx.doi.org/10.1007/JHEP01(2013)080}{JHEP
  {\bfseries 1301} (2013) 080},
\href{http://arxiv.org/abs/1210.6832}{{\ttfamily arXiv:1210.6832 [hep-ph]}}.

\bibitem{Czakon:2013goa}
M.~Czakon, P.~Fiedler, and A.~Mitov,
  \href{http://dx.doi.org/10.1103/PhysRevLett.110.252004}{Phys. Rev. Lett.
  {\bfseries 110} (2013) 252004},
\href{http://arxiv.org/abs/1303.6254}{{\ttfamily arXiv:1303.6254 [hep-ph]}}.

\bibitem{TopPP:2011}
M.~Czakon and A.~Mitov,
\href{http://arxiv.org/abs/1112.5675}{{\ttfamily arXiv:1112.5675 [hep-ph]}}.

\bibitem{Botje:2011sn}
M.~Botje {et~al.},
\href{http://arxiv.org/abs/1101.0538}{{\ttfamily arXiv:1101.0538 [hep-ph]}}.

\bibitem{mstw_alphas}
A.~D. Martin, W.~J. Stirling, R.~S. Thorne, and G.~Watt,
  \href{http://dx.doi.org/10.1140/epjc/s10052-009-1164-2}{Eur. Phys. J.
  {\bfseries C64} (2009) 653},
\href{http://arxiv.org/abs/0905.3531}{{\ttfamily arXiv:0905.3531 [hep-ph]}}.

\bibitem{Gao:2013xoa}
J.~Gao {et~al.}, \href{http://dx.doi.org/10.1103/PhysRevD.89.033009}{Phys. Rev.
  {\bfseries D89} (2014) 033009},
\href{http://arxiv.org/abs/1302.6246}{{\ttfamily arXiv:1302.6246 [hep-ph]}}.

\bibitem{nnpdf23}
R.~Ball {et~al.},
  \href{http://dx.doi.org/10.1016/j.nuclphysb.2012.10.003}{Nucl. Phys.
  {\bfseries B867} (2013) 244},
\href{http://arxiv.org/abs/1207.1303}{{\ttfamily arXiv:1207.1303 [hep-ph]}}.

\bibitem{Kidonakis:2010ux}
N.~Kidonakis, \href{http://dx.doi.org/10.1103/PhysRevD.82.054018}{Phys. Rev.
  {\bfseries D82} (2010) 054018},
\href{http://arxiv.org/abs/1005.4451}{{\ttfamily arXiv:1005.4451 [hep-ph]}}.

\bibitem{calchep}
A.~Belyaev, N.~D. Christensen, and A.~Pukhov,
  \href{http://dx.doi.org/10.1016/j.cpc.2013.01.014}{Comput. Phys. Commun.
  {\bfseries 184} (2013) 1729},
  \href{http://arxiv.org/abs/1207.6082}{{\ttfamily arXiv:1207.6082 [hep-ph]}}.

\bibitem{ref:Gingrich_generator}
D.~M. Gingrich, \href{http://dx.doi.org/10.1016/j.cpc.2010.07.027}{Comput.
  Phys. Commun. {\bfseries 181} (2010) 1917},
  \href{http://arxiv.org/abs/0911.5370}{{\ttfamily arXiv:0911.5370 [hep-ph]}}.
  \url{http://qbh.hepforge.org/}.

\bibitem{Mathews:2004xp}
P.~Mathews, V.~Ravindran, K.~Sridhar, and W.~L. van Neerven,
  \href{http://dx.doi.org/10.1016/j.nuclphysb.2005.01.051}{Nucl. Phys.
  {\bfseries B713} (2005) 333},
  \href{http://arxiv.org/abs/hep-ph/0411018}{{\ttfamily arXiv:hep-ph/0411018}}.

\bibitem{Mathews:2005bw}
P.~Mathews, V.~Ravindran, and K.~Sridhar,
  \href{http://dx.doi.org/10.1088/1126-6708/2005/10/031}{JHEP {\bfseries 0510}
  (2005) 031}, \href{http://arxiv.org/abs/hep-ph/0506158}{{\ttfamily
  arXiv:hep-ph/0506158}}.

\bibitem{Kumar:2006id}
M.~C. Kumar, P.~Mathews, and V.~Ravindran,
  \href{http://dx.doi.org/10.1140/epjc/s10052-006-0054-0}{Eur. Phys. J.
  {\bfseries C49} (2007) 599},
  \href{http://arxiv.org/abs/hep-ph/0604135}{{\ttfamily arXiv:hep-ph/0604135}}.

\bibitem{atlas:egamma_perf}
{{ATLAS}} Collaboration,
  \href{http://dx.doi.org/10.1140/epjc/s10052-012-1909-1}{Eur.~Phys.~J.
  {\bfseries C72} (2012) 1909},
\href{http://arxiv.org/abs/1110.3174}{{\ttfamily arXiv:1110.3174 [hep-ex]}}.

\bibitem{Aad:2014zya}
{ATLAS} Collaboration,
\href{http://arxiv.org/abs/1404.4562}{{\ttfamily arXiv:1404.4562 [hep-ex]}}.

\bibitem{muon_reco_paper}
{ATLAS Collaboration}, ATLAS-CONF-2013-088 (2013).
  \url{http://cds.cern.ch/record/1580207}.

\bibitem{Aad:2011eps}
{ATLAS Collaboration},
  \href{http://dx.doi.org/10.1103/PhysRevLett.107.272002}{Phys.~Rev.~Lett.
  {\bfseries 107} (2011) 272002},
  \href{http://arxiv.org/abs/1108.1582}{{\ttfamily arXiv:1108.1582 [hep-ex]}}.

\bibitem{vrap}
C.~Anastasiou, L.~Dixon, K.~Melnikov, and F.~Petriello,
  \href{http://dx.doi.org/10.1103/PhysRevD.69.094008}{Phys.~Rev.~D {\bfseries
  69} (2004) 094008}, \href{http://arxiv.org/abs/hep-ph/0312266}{{\ttfamily
  arXiv:hep-ph/0312266}}.

\bibitem{Alekhin:2012ig}
{S. Alekhin, J. Blumlein and S. Moch},
  \href{http://dx.doi.org/10.1103/PhysRevD.86.054009}{Phys. Rev. {\bfseries
  D86} (2012) 054009}, \href{http://arxiv.org/abs/1202.2281}{{\ttfamily
  arXiv:1202.2281 [hep-ph]}}.

\bibitem{herapdf10}
{H1 and ZEUS Collaborations, F.D. Aaron et al.},
  \href{http://dx.doi.org/10.1007/JHEP01(2010)109}{JHEP {\bfseries 1001} (2010)
  109}, \href{http://arxiv.org/abs/0911.0884}{{\ttfamily arXiv:0911.0884
  [hep-ex]}}.

\bibitem{beam_energy}
J.~Wenninger, CERN-ATS-2013-040 (2013).
  \url{http://cds.cern.ch/record/1546734}.

\bibitem{Aad:2011xp}
{ATLAS Collaboration},
  \href{http://dx.doi.org/10.1016/j.physletb.2011.04.044}{Phys.~Lett.
  {\bfseries B700} (2011) 163},
  \href{http://arxiv.org/abs/1103.6218}{{\ttfamily arXiv:1103.6218 [hep-ex]}}.

\bibitem{CDF:Zpmumu2fb}
{CDF Collaboration, T. Aaltonen et al.},
  \href{http://dx.doi.org/10.1103/PhysRevLett.102.091805}{Phys.~Rev.~Lett.
  {\bfseries 102} (2009) 091805},
\href{http://arxiv.org/abs/0811.0053}{{\ttfamily arXiv:0811.0053 [hep-ex]}}.

\bibitem{bayesianMethod}
A.~Caldwell, D.~Kollar, and K.~Kr{\"o}ninger,
  \href{http://dx.doi.org/10.1016/j.cpc.2009.06.026}{Comput.~Phys.~Commun.
  {\bfseries 180} (2009) 2197},
  \href{http://arxiv.org/abs/0808.2552}{{\ttfamily arXiv:0808.2552
  [physics.data-an]}}.

\end{thebibliography}\endgroup

\onecolumngrid
\clearpage
\begin{flushleft}
{\Large The ATLAS Collaboration}

\bigskip

G.~Aad$^{\rm 84}$,
B.~Abbott$^{\rm 112}$,
J.~Abdallah$^{\rm 152}$,
S.~Abdel~Khalek$^{\rm 116}$,
O.~Abdinov$^{\rm 11}$,
R.~Aben$^{\rm 106}$,
B.~Abi$^{\rm 113}$,
M.~Abolins$^{\rm 89}$,
O.S.~AbouZeid$^{\rm 159}$,
H.~Abramowicz$^{\rm 154}$
% ...
\bigskip
\\
% ...
$^{11}$ Institute of Physics, Azerbaijan Academy of Sciences, Baku, Azerbaijan\\
% ...
$^{84}$ CPPM, Aix-Marseille Universit{\'e} and CNRS/IN2P3, Marseille, France\\
% ...
$^{89}$ Department of Physics and Astronomy, Michigan State University, East Lansing MI, United States of America\\
% ...
$^{106}$ Nikhef National Institute for Subatomic Physics and University of Amsterdam, Amsterdam, Netherlands\\
% ...
$^{112}$ Homer L. Dodge Department of Physics and Astronomy, University of Oklahoma, Norman OK, United States of America\\
$^{113}$ Department of Physics, Oklahoma State University, Stillwater OK, United States of America\\
% ...
$^{116}$ LAL, Universit{\'e} Paris-Sud and CNRS/IN2P3, Orsay, France\\
% ...
$^{152}$ Institute of Physics, Academia Sinica, Taipei, Taiwan\\
% ...
$^{154}$ Raymond and Beverly Sackler School of Physics and Astronomy, Tel Aviv University, Tel Aviv, Israel\\
% ...
$^{159}$ Department of Physics, University of Toronto, Toronto ON, Canada\\
% ...
\end{flushleft}

\end{document}